\documentclass[acmsmall]{acmart}

\setcopyright{acmlicensed}
\acmJournal{PACMHCI}
\acmYear{2021} \acmVolume{5} \acmNumber{CSCW2} \acmArticle{386} \acmMonth{10} \acmPrice{15.00}\acmDOI{10.1145/3479530}

\AtBeginDocument{%
  \providecommand\BibTeX{{%
    \normalfont B\kern-0.5em{\scshape i\kern-0.25em b}\kern-0.8em\TeX}}}

\usepackage{lookup}

\lookupPut{M1_01}{1}
\lookupPut{M1_02}{2}
\lookupPut{M1_03}{3}
\lookupPut{M1_04}{4}
\lookupPut{M1_05}{5}
\lookupPut{M1_06}{6}
\lookupPut{M1_07}{7}
\lookupPut{M1_08}{8}
\lookupPut{M1_09}{9}
\lookupPut{M1_10}{10}
\lookupPut{M1_11}{11}
\lookupPut{M1_13}{12}
\lookupPut{M1_14}{13}
\lookupPut{M1_15}{14}
\lookupPut{M1_16}{15}
\lookupPut{M1_17}{16}
\lookupPut{M1_19}{17}
\lookupPut{M1_20}{18}
\lookupPut{M1_21}{19}
\lookupPut{M1_22}{20}
\lookupPut{M1_23}{21}
\lookupPut{M1_24}{22}
\lookupPut{M1_25}{23}
\lookupPut{M1_26}{24}
\lookupPut{M1_27}{25}
\lookupPut{M1_28}{26}
\lookupPut{M1_29}{27}
\lookupPut{M1_30}{28}
\lookupPut{M1_31}{29}
\lookupPut{M1_33}{30}
\lookupPut{M1_34}{31}
\lookupPut{M1_37}{32}
\lookupPut{M1_38}{33}
\lookupPut{M1_39}{34}
\lookupPut{M1_40}{35}
\lookupPut{M1_42}{36}
\lookupPut{M1_46}{37}
\lookupPut{M1_47}{38}
\lookupPut{M1_48}{39}
\lookupPut{M1_49}{40}
\lookupPut{M1_50}{41}
\lookupPut{M1_51}{42}
\lookupPut{M1_52}{43}
\lookupPut{M1_54}{44}
\lookupPut{M1_55}{45}
\lookupPut{M1_56}{46}
\lookupPut{M1_57}{47}
\lookupPut{M1_58}{48}
\lookupPut{M1_60}{49}
\lookupPut{M1_61}{50}
\lookupPut{M1_63}{51}
\lookupPut{M1_64}{52}
\lookupPut{M1_65}{53}
\lookupPut{M1_66}{54}
\lookupPut{M1_67}{55}
\lookupPut{M1_68}{56}
\lookupPut{M1_69}{57}
\lookupPut{M1_70}{58}
\lookupPut{M1_71}{59}
\lookupPut{M1_72}{60}
\lookupPut{M1_74}{61}
\lookupPut{M1_75}{62}
\lookupPut{M1_78}{63}
\lookupPut{M1_82}{64}
\lookupPut{M1_84}{65}
\lookupPut{M1_85}{66}
\lookupPut{M1_86}{67}
\lookupPut{M1_87}{68}
\lookupPut{M1_88}{69}
\lookupPut{M1_91}{70}
\lookupPut{M1_92}{71}
\lookupPut{M1_93}{72}
\lookupPut{M2_94}{73}
\lookupPut{M2_95}{74}
\lookupPut{M2_96}{75}
\lookupPut{M2_98}{76}
\lookupPut{M2_99}{77}
\lookupPut{M2_100}{78}
\lookupPut{M2_101}{79}
\lookupPut{M2_102}{80}
\lookupPut{M2_103}{81}
\lookupPut{M2_104}{82}
\lookupPut{M2_105}{83}
\lookupPut{M2_106}{84}
\lookupPut{M2_107}{85}
\lookupPut{M2_110}{86}
\lookupPut{M2_111}{87}
\lookupPut{M2_112}{88}
\lookupPut{M2_113}{89}
\lookupPut{M2_114}{90}
\lookupPut{M2_115}{91}
\lookupPut{M2_118}{92}
\lookupPut{M2_119}{93}
\lookupPut{M2_120}{94}
\lookupPut{M2_121}{95}
\lookupPut{M2_122}{96}
\lookupPut{M2_123}{97}
\lookupPut{M2_124}{98}
\lookupPut{M2_125}{99}
\lookupPut{M2_126}{100}
\lookupPut{M2_127}{101}
\lookupPut{M2_128}{102}
\lookupPut{M2_129}{103}
\lookupPut{M2_130}{104}
\lookupPut{M2_131}{105}
\lookupPut{M2_132}{106}
\lookupPut{M2_133}{107}
\lookupPut{M2_134}{108}
\lookupPut{M3_135}{109}
\lookupPut{M3_136}{110}
\lookupPut{M3_137}{111}
\lookupPut{M3_138}{112}
\lookupPut{M3_139}{113}
\lookupPut{M3_141}{114}
\lookupPut{M3_142}{115}
\lookupPut{M4_144}{116}
\lookupPut{M4_145}{117}
\lookupPut{M4_146}{118}
\lookupPut{M5_149}{119}
\lookupPut{M5_150}{120}
\lookupPut{M5_152}{121}
\lookupPut{M5_153}{122}
\lookupPut{MS_16}{123}
\lookupPut{MS_17}{124}
\lookupPut{MS_18}{125}
\lookupPut{MS_19}{126}
\lookupPut{MS_2}{127}
\lookupPut{MS_7}{128}
\lookupPut{MS_8}{129}
\lookupPut{MS_9}{130}

\lookupPut{M1}{1}
\lookupPut{M2}{2}
\lookupPut{M3}{3}
\lookupPut{M4}{4}
\lookupPut{M5}{5}
\lookupPut{M6}{6}
\lookupPut{M7}{7}
\lookupPut{M8}{8}
\lookupPut{M9}{9}
\lookupPut{M10}{10}
\lookupPut{M11}{11}
\lookupPut{M13}{12}
\lookupPut{M14}{13}
\lookupPut{M15}{14}
\lookupPut{M16}{15}
\lookupPut{M17}{16}
\lookupPut{M19}{17}
\lookupPut{M20}{18}
\lookupPut{M21}{19}
\lookupPut{M22}{20}
\lookupPut{M23}{21}
\lookupPut{M24}{22}
\lookupPut{M25}{23}
\lookupPut{M26}{24}
\lookupPut{M27}{25}
\lookupPut{M28}{26}
\lookupPut{M29}{27}
\lookupPut{M30}{28}
\lookupPut{M31}{29}
\lookupPut{M33}{30}
\lookupPut{M34}{31}
\lookupPut{M37}{32}
\lookupPut{M38}{33}
\lookupPut{M39}{34}
\lookupPut{M40}{35}
\lookupPut{M42}{36}
\lookupPut{M46}{37}
\lookupPut{M47}{38}
\lookupPut{M48}{39}
\lookupPut{M49}{40}
\lookupPut{M50}{41}
\lookupPut{M51}{42}
\lookupPut{M52}{43}
\lookupPut{M54}{44}
\lookupPut{M55}{45}
\lookupPut{M56}{46}
\lookupPut{M57}{47}
\lookupPut{M58}{48}
\lookupPut{M60}{49}
\lookupPut{M61}{50}
\lookupPut{M63}{51}
\lookupPut{M64}{52}
\lookupPut{M65}{53}
\lookupPut{M66}{54}
\lookupPut{M67}{55}
\lookupPut{M68}{56}
\lookupPut{M69}{57}
\lookupPut{M70}{58}
\lookupPut{M71}{59}
\lookupPut{M72}{60}
\lookupPut{M74}{61}
\lookupPut{M75}{62}
\lookupPut{M78}{63}
\lookupPut{M82}{64}
\lookupPut{M84}{65}
\lookupPut{M85}{66}
\lookupPut{M86}{67}
\lookupPut{M87}{68}
\lookupPut{M88}{69}
\lookupPut{M91}{70}
\lookupPut{M92}{71}
\lookupPut{M93}{72}
\lookupPut{M94}{73}
\lookupPut{M95}{74}
\lookupPut{M96}{75}
\lookupPut{M98}{76}
\lookupPut{M99}{77}
\lookupPut{M100}{78}
\lookupPut{M101}{79}
\lookupPut{M102}{80}
\lookupPut{M103}{81}
\lookupPut{M104}{82}
\lookupPut{M105}{83}
\lookupPut{M106}{84}
\lookupPut{M107}{85}
\lookupPut{M110}{86}
\lookupPut{M111}{87}
\lookupPut{M112}{88}
\lookupPut{M113}{89}
\lookupPut{M114}{90}
\lookupPut{M115}{91}
\lookupPut{M118}{92}
\lookupPut{M119}{93}
\lookupPut{M120}{94}
\lookupPut{M121}{95}
\lookupPut{M122}{96}
\lookupPut{M123}{97}
\lookupPut{M124}{98}
\lookupPut{M125}{99}
\lookupPut{M126}{100}
\lookupPut{M127}{101}
\lookupPut{M128}{102}
\lookupPut{M129}{103}
\lookupPut{M130}{104}
\lookupPut{M131}{105}
\lookupPut{M132}{106}
\lookupPut{M133}{107}
\lookupPut{M134}{108}
\lookupPut{M135}{109}
\lookupPut{M136}{110}
\lookupPut{M137}{111}
\lookupPut{M138}{112}
\lookupPut{M139}{113}
\lookupPut{M141}{114}
\lookupPut{M142}{115}
\lookupPut{M144}{116}
\lookupPut{M145}{117}
\lookupPut{M146}{118}
\lookupPut{M149}{119}
\lookupPut{M150}{120}
\lookupPut{M152}{121}
\lookupPut{M153}{122}

\lookupPut{MS16}{123}
\lookupPut{MS17}{124}
\lookupPut{MS18}{125}
\lookupPut{MS19}{126}
\lookupPut{MS2}{127}
\lookupPut{MS7}{128}
\lookupPut{MS8}{129}
\lookupPut{MS9}{130}

\lookupPut{P02}{01}
\lookupPut{P04}{02}
\lookupPut{P05}{03}
\lookupPut{P06}{04}
\lookupPut{P08}{05}
\lookupPut{P11}{06}
\lookupPut{P12}{07}
\lookupPut{P13}{08}
\lookupPut{P14}{09}
\lookupPut{P16}{10}
\lookupPut{P17}{11}
\lookupPut{P18}{12}
\lookupPut{P20}{13}
\lookupPut{P21}{14}
\lookupPut{P22}{15}
\lookupPut{P23}{16}
\lookupPut{I02}{01}
\lookupPut{I04}{02}
\lookupPut{I05}{03}
\lookupPut{I06}{04}
\lookupPut{I08}{05}
\lookupPut{I11}{06}
\lookupPut{I12}{07}
\lookupPut{I13}{08}
\lookupPut{I14}{09}
\lookupPut{I16}{10}
\lookupPut{I17}{11}
\lookupPut{I18}{12}
\lookupPut{I20}{13}
\lookupPut{I21}{14}
\lookupPut{I22}{15}
\lookupPut{I23}{16}
\lookupPut{V02}{01}
\lookupPut{V04}{02}
\lookupPut{V05}{03}
\lookupPut{V06}{04}
\lookupPut{V08}{05}
\lookupPut{V11}{06}
\lookupPut{V12}{07}
\lookupPut{V13}{08}
\lookupPut{V14}{09}
\lookupPut{V16}{10}
\lookupPut{V17}{11}
\lookupPut{V18}{12}
\lookupPut{V20}{13}
\lookupPut{V21}{14}
\lookupPut{V22}{15}
\lookupPut{V23}{16}

\usepackage{tabularx}
\usepackage{amsmath}
\usepackage{amssymb}
\usepackage{amsfonts}
\usepackage{algpseudocode,algorithm,algorithmicx} 
\usepackage{graphicx}
\usepackage{textcomp}
\usepackage[ampersand]{easylist}
\usepackage{multicol}
\usepackage{multirow}

\usepackage{threeparttable}
\usepackage{tabularx}

\usepackage{rotating}
\usepackage[roman]{parnotes}
\usepackage[many]{tcolorbox}
\usepackage[font={small}]{caption}
\usepackage{soul}
\usepackage{xcolor}
\usepackage{comment}
\usepackage{subcaption}
\usepackage{booktabs}
\usepackage{colortbl}
\usepackage{verbatim}

\usepackage{xspace}
\xspaceaddexceptions{]\}}

\usepackage{hyperref}
\usepackage{array}
\usepackage[inline]{enumitem}
\usepackage[export]{adjustbox}
\usepackage{longfbox}
\usepackage{commath}
\usepackage[normalem]{ulem}
\usepackage{mdframed,lipsum}
\usepackage{ifdraft}

\usepackage{makecell}

\usepackage{fontawesome}

\usepackage{xcolor} %
\definecolor{darkgreen}{rgb}{0.05,0.5,0.05}

\newcommand{\yt}{YouTube\xspace}

\usepackage{framed}
\usepackage[strict]{changepage}

\definecolor{PAblue}{RGB}{0,122,204}%

\newcommand{\quo}[2]{ 
	\vspace{-0.15cm}
	\def\FrameCommand{%
		\hspace{0pt}%
		{\color{PAblue}\vrule width 2pt}%
		{\color{white}\vrule width 2pt}%
		\colorbox{white}
	}%
	\MakeFramed{\advance\hsize-\width\FrameRestore}%
	\noindent\hspace{-4.55pt}%
	\begin{adjustwidth}{}{0pt}
		\vspace{-2pt}%
		``\emph{#1}'' ({#2})
		\vspace{-3pt}
	\end{adjustwidth}\endMakeFramed%
	\vspace{-0.15cm}
}

\newcommand{\inlinequote}[1]{``#1''}

\newcommand{\RQA}{What are developer motivations and intentions for creating `day in the life' vlogs?\xspace}
\newcommand{\RQB}{What content do developers share?\xspace}
\newcommand{\RQC}{What kind of interaction happens around these videos? How are these videos received by the broader community?\xspace}

\DeclareRobustCommand{\mybox}[2][gray!13]{%
\begin{tcolorbox}[   %
        breakable,
        left=1pt,
        right=1pt,
        top=1pt,
        bottom=1pt,
        colback=#1,
        colframe=black,
        width=\dimexpr\columnwidth\relax, 
        enlarge left by=0mm,
        boxsep=5pt,
        ]
        #2
\end{tcolorbox}
}

\definecolor{lightgreen}{RGB}{247,253,251}
\definecolor{newgreen}{RGB}{7,94,70}
\definecolor{bordergreen}{RGB}{101,223,190}

\newtcolorbox{myideabox}{sidebyside,
                         colback=lightgreen,
                         colframe=bordergreen, 
                         coltext=newgreen,
                         boxrule=0.9pt,
                         boxsep=5pt,
                         arc=1pt,
                         leftrule=0.8mm, 
                         lefthand width=0.4cm, 
                         lower separated=false, 
                         sidebyside gap=3mm, 
                                 right=1pt,
                            top=1pt,
        bottom=1pt,
                         left=3pt}

\newcounter{takeaway}%

\DeclareRobustCommand{\takeawayboxx}[2]{\refstepcounter{takeaway}\label{#1}%
\begin{myideabox}\centering\LARGE\faLightbulbO\tcblower
\textbf{Finding \thetakeaway:} #2
\end{myideabox}
}

\RequirePackage{color}
\definecolor{BLUE}{rgb}{0,0,1}
\definecolor{BLACK}{rgb}{0,0,0}

\newcommand{\BEGINADDED}{\protect\color{BLACK}}
\newcommand{\BEGINADDEDX}{\color{BLACK}}

\newcommand{\ENDADDED}{\protect\color{black}}

\begin{document}

\title{Developers Who Vlog: Dismantling Stereotypes through Community and Identity}

\author{Souti Chattopadhyay}
\email{chattops@oregonstate.edu}
\affiliation{%
  \institution{Oregon State University}
  \city{Corvallis}
  \state{Oregon}
  \country{USA}
}

\author{Denae Ford}
\email{denae@microsoft.com}
\affiliation{%
  \institution{Microsoft Research}
  \city{Redmond}
  \state{Washington}
  \country{USA}
}

\author{Thomas Zimmermann}
\email{tzimmer@microsoft.com}
\affiliation{%
  \institution{Microsoft Research}
  \city{Redmond}
  \state{Washington}
  \country{USA}
}

\date{October 2021}

\renewcommand{\shortauthors}{Souti Chattopadhyay, Denae Ford, \& Thomas Zimmermann}

\begin{abstract}

Developers are more than ``nerds behind computers all day'', they lead a normal life, and not all take the traditional path to learn programming. However, the public still sees software development as a profession for ``math wizards''. 
To learn more about this special type of knowledge worker from their first-person perspective, we conducted three studies to learn how developers describe a day in their life through vlogs on YouTube and how these vlogs were received by the broader community.
We first interviewed 16 developers who vlogged to identify their motivations for creating this content and their intention behind what they chose to portray. Second, we analyzed 130 vlogs (video blogs) to understand the range of the content conveyed through videos. Third, we analyzed 1176 comments from the 130 vlogs to understand the impact the vlogs have on the audience. 
We found that developers were motivated to promote and build a diverse community, by sharing different aspects of life that define their identity, and by creating awareness about learning and career opportunities in computing.
They used vlogs to share a variety of how software developers work and live---showcasing often unseen experiences, including intimate moments from their personal life.
From our comment analysis, we found that the vlogs were valuable to the audience to find information and seek advice. Commenters sought opportunities to connect with others over shared triumphs and trials they faced that were also shown in the vlogs. 
As a central theme, we found that developers use vlogs to challenge the misconceptions and stereotypes around their identity, work-life, and well-being. These social stigmas are obstacles to an inclusive and accepting community and can deter people from choosing software development as a career. 
We also discuss the implications of using vlogs to support developers, researchers, and beyond.
\end{abstract}



\maketitle

\section{Introduction}

Vlogs, or video blogs, have been a powerful tool for sharing personal experiences online. They are often used as vessel for understanding experiences of individuals on a day to day that are not as easily showcased in text. For example, explaining complex collaborative work such as software development can be made easier to explain by showing others rather than communicating via a written blog. Beyond explaining a technical concept, vlogs can be an insert into the full experiences that people face---as an alternative to physically shadowing a person all day. In these settings, shadowers can get deeper understanding for the entire role, which is especially important for the increasingly high-demand career of software development~\cite{columbus2020mostindemand}.

As Software developers, being a special type of knowledge worker~\cite{meyer2017design}, often find themselves on the cutting edge of technology systems, they often build tools to support their work that can later support other types of knowledge workers in technology (e.g., Emails, Internet Search, and Wiki Pages)~\cite{kelly2014prototype}. For this reason, many have studied them as the prototype of knowledge workers---often pushing the boundaries of knowledge work~\cite{kelly2008changing}.

\begin{figure*}[!tbp]
\centering
\includegraphics[width=\textwidth]{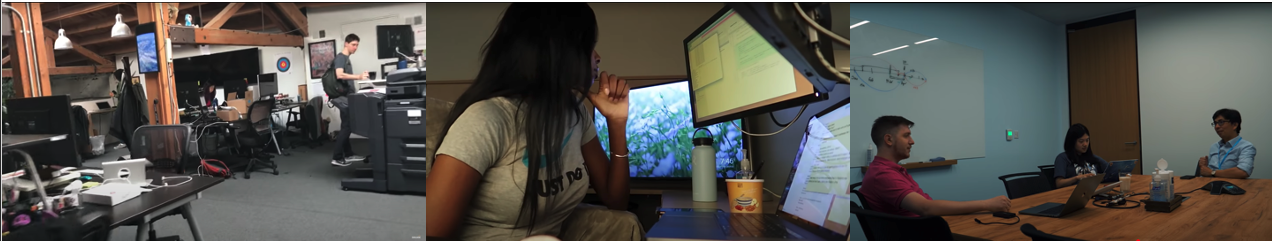}

\vspace{-6pt}
\caption{\BEGINADDED Vlogs capture an intimate view of a developer's life showing where and \textbf{how developers work} how often they interact with their coworkers e.g. developers work in close proximity and seamlessly collaborate (left), some developers work in their own offices (center), or other activities at work like meetings (right). These vlogs give a full picture of a developer's day.}
\label{fig:vlog_at_work1}
\end{figure*}

Although, what it means to be a software developers has been studied, much of their experiences have been limited to reported estimations of time or retrospective interviews~\cite{meyer2019today} which are not the most reflective of what it fully means to take up this career. There has been limited accounts of the day-to-day experiences where developers can control the narrative of what they decide is important to know about this technical career.

\BEGINADDEDX
Our motivation to study developers’ vlogs comes from (1) the need for more research to understand the different identities and lifestyles of developers to create work environments to support these diverse populations of developers, and also (2) to investigate the perception of the general public, from both technical and non-technical background, about the software development industry in general as it impacts how we can support and encourage the future generation of developers. Vlogs provide a unique source of data for this investigation, providing a deeper look into the developers’ experiences where they control the narrative themselves. Thus, vlogs give a first person perspective of what activities and lifestyle related factors of a developers’ life they consider important.
\ENDADDED

\begin{figure*}[!tbp]
\includegraphics[width=\textwidth]{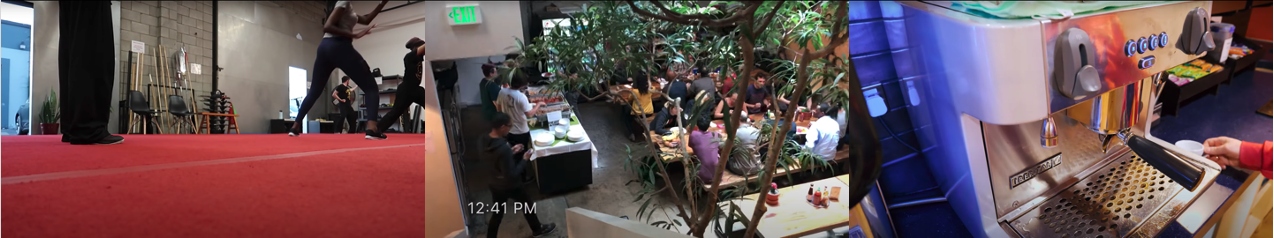}

\vspace{-6pt}
\caption{\BEGINADDED Vlogs also give us a glimpse of \textbf{what developers do outside work} e.g., practicing Taekwondo (left), how they spend their social hours e.g., developer grabbing lunch with friends (center), or their me-time at work e.g., developer brewing a cappuccino for themselves (right). The vlogs give us an idea about the developer's lifestyles and the hobbies they enjoy.}
\label{fig:vlog_at_work2}
\end{figure*}

\BEGINADDED Over the past years, many developers have described a day in their life through vlogs on \yt.%
\footnote{\BEGINADDED To get an idea of the content of these vlogs, we recommend the reader to watch the following examples: \begin{enumerate} \item ``Day in the Life of a Software Engineer (First week!)'' (505K views),  \url{https://youtu.be/bX8hvldRx1M}; \item ``a day in the life of a software engineer'' (5 million views),  \url{https://youtu.be/rqX8PFcOpxA}; and \item ``Day of Amazon Software Developer'' (519K views), \url{https://youtu.be/c8dd9f5MamU} \end{enumerate}}
These vlogs show how developers work (see Figure~\ref{fig:vlog_at_work1}) and also what they do outside of work (see Figure~\ref{fig:vlog_at_work2}). \ENDADDED
Given the diverse personal narratives presented in vlogs and the autonomy left to the creator, these videos give developers a platform to shine a light on the detailed experience of what it means to be a software developer—of which can also dismantle stereotypes of what people think software developers do.  For instance, there are several negative stereotypes of what a software developer ``should’’ be such as working in isolation from others~\cite{quora2019stereotypes}, having a formal university degree with strong math skills~\cite{codingdojo2020myths}, and identifying as nonathletic, white male~\cite{pantoja2017stereotypes}. 
This stigma can discourage the emerging generation of software developers who are interested in pursuing what can actually be a collaborative and social career~\cite{rochester2016stereotyping}.
Although these negative perceptions exist, developers have been doing the work to strategically dismantle these stereotypes through the use of hashtag movements~\cite{liu2017selfies}, written blog posts~\cite{chiang2020transition}, and podcasts~\cite{codenewbiespodcast}. However, there has been limited research on understanding how creators and consumers of vlogs can combat these misconceptions by showcasing experiences where the technical workers can control their own story of their profession. %

In this paper, we investigate the full scope of software developer experiences through vlogs. \BEGINADDED We study how developers describe a day in their life through vlogs on YouTube and how these vlogs were received by the broader community.
Specifically, we conducted a qualitative study to better understand the types of content is included in these videos, motivations behind why software developers vlog, \ENDADDED and the perceived impact from these vlogs. We chose to study software developers who vlog because social media, such as Twitter~\cite{liu2017selfies, fang2020need} and blogs~\cite{storey2014revolution} continue to be a growing resource software developers use to discuss their technical work which intersects with their personal experiences. Likewise platforms, like \yt where those interested in programming can find tutorials on programming concepts~\cite{parra2018automatic}, are now also using this platform as a place to be transparent about what it is like to be a software developer in its entirety. Our study dives deep to understand the types of experiences shared by conducting a content analysis on 130 ‘day in the life of a software developer’ vlogs from 21 countries and interviewing 16 vloggers about their motivations for sharing the content they present. 
Likewise, to better understand the impact of these vlogs, we conducted a thematic analysis of comments to find how other developers and other viewers found value in the topics presented.

We found that vlogs showcased a deep set of experiences at the intersection of developers' professional and personal life---a full perspective that is difficult to capture through traditional data collection such as interviews and surveys.  
By not only telling but by showing, vlogs are able to cover topics around company culture, work-life balance, and interactions with colleagues.
In terms of their roles, many empirical studies overly simplify characterizing software developers as being a ``math wizard'' from a university or a ``no time for fun'' workaholic at a large technology company. 
However, through this archive of first-person perspectives, vlogs show that developers assume a variety of collaborative roles and approaches to their careers. 
Some of this includes freelancing, being a lead developer, working at a startup, or  transitioning from a music career and learning to code as they go.
We found that through these videos, vloggers were able to create a community of emerging and existing software developers that were both inspired to enter and continue careers in software development. Vlogs provided developers an opportunity to be transparent about their career, its transitions (such as losing a job), and the journey through while also providing a space for others to share similar challenges faced and provide support for one another.
In summary, this work contributes to the body of literature the understanding of the full experiences of how technical workers create awareness and inspire others to pursue careers by:
\begin{enumerate}
\item providing empirical findings on the motivations, content, and perceived impact of software developer vlogs (Sections~\ref{sec:study-interviews}, \ref{sec:study-vloganalysis}, and~\ref{sec:study-comments}),
\item providing a discussion of several software developer stereotypes (Section~\ref{sec:stereotypes}), 
\item \BEGINADDED outlining best practices and implications for content creators, video sharing platforms, and research (Sections~\ref{sec:developers} and~\ref{sec:research-implications}).\ENDADDED
\end{enumerate}

\section{Related Work}

\BEGINADDED \yt is a video platform that hosts more than 26 billion videos covering topics from entertainment to education. The people involved in creating these videos, the viewers and their interaction with the content, and the range of content available makes \yt an interesting subject for researchers. In the past decade, researchers have studied the structure and working of \yt's different components like recommendation algorithms, views and comment, with the intention of improving the features through automation and ML ~\cite{paolillo2008core, poche2017comments}. 
Others have studied \yt from the social lens, how \yt can generate open, authentic communities with participatory culture~\cite{burgess2018youtube} and the effect on these communities in different directions like benefits and challenges of using \yt for education~\cite{jones2011youtube, carlisle2010using}, effect of social networking among youth~\cite{lange2007publicly}, impact on propagating healthcare information~\cite{keelan2007youtube, madathil2015healthcare} etc. 

However, the impact of \yt videos on the software engineering community has largely been explored based on technical videos curated for educations purposes. \ENDADDED Online videos have been helpful for learning new technical content. For example, technical content being available as interactive lessons through MOOCs have been helpful with teaching technical content in large course settings~\cite{christensen2013mooc,hadavand2018can}. However, MOOCS often leave out the applied experience of what it means to us what was learned in practice. More specifically, \emph{how} it will be in practice or \emph{if} it will at all. In fact the low completion rates of MOOCs demonstrate that they are transferring into increased labor market value~\cite{lederman2019moocs}.
\BEGINADDED However, these online videos are missing  the practicality of what is being shared, creating a lack of confidence among viewers regarding the applicability. \ENDADDED

\subsection{Vlogs and their unique features}

\BEGINADDED Vlogs, a subset of the types of videos on \yt, are unique for bringing forth the sense of practicality by sharing experiences in first person. In this paper, we study the effect of vlogs on \yt generated by developers from the software engineering community. As Fidan et al.~\cite{fidan2018usage} put it ``the most significant feature of vlogs is that a person forms his/her own experiences and share them as videos.'' \ENDADDED
To find practical experiences at scale, people often turn to sharing vlogs on platforms like, \yt. \yt provides a good fit to share content by allowing creators being able to upload videos for free, the autonomy on the content of the video, and the ability for viewers to engage with content \BEGINADDED as well as other viewers who have shared experiences similar or contrasting to the vloggers.\ENDADDED

Vlogs can be used as a tool to share and spread unique experiences. For example, among other communities, creators have used vlogs to connect people with chronic diseases~\cite{liu2013health}, served as a public vehicle for gender transitions~\cite{raun2015video}, help resources for parents supporting children with developmental disabilities~\cite{rodriguez2019myautsome}, or document program knowledge in the form of screencasts~\cite{macleod2015code,ponzanelli2016too,yadid2016extracting}. In learning day-to-day life experiences, vlogs can be a helpful way to get a front row seat at the in-depth experiences people face. \BEGINADDED These studies reveal features specific to these kinds of vlogs like ``methods that health vloggers use to establish a connection with their viewers like \ldots health vloggers explicitly sought interaction with their viewers'' as health related issues generate a strong emotional response~\cite{liu2013health}. However, we don't know why the developer community creates vlogs, and the information and experiences they share with their audiences. In this paper, we study vlogs from developers to understand their motivation, the content of the vlogs and the type of connection they build with their audience. \ENDADDED 

Although most vlogs are uniquely positioned for there to be one person broadcasting their experiences and not engaging in synchronous engagement as tasks are going on, there are still opportunities for others to learn by peripheral observation and asynchronous engagement with peers. In the case of developer vlogs, each vlog can be considered a \emph{community of practice} where through comments interested community members can describe how insightful the experiences shared in the video were to their personal journey.

\subsection{Roles and Trajectories of Software Developers}
The trajectory and practical impact of people who build software have been studied as they evolve from junior to senior positions~\cite{garden1990career} and as they progress in leaderships roles across OSS projects~\cite{trinkenreich2020hidden}. While this gives us an overview of developers' career and journey, to learn the successful pathways of transitions across roles we need to understand their routines and everyday practices.
In order to understand developers in a deeper approach there have also been studies to understand how developer spend their time in a day~\cite{ford2017characterizing} and the technical tasks they focus on ~\cite{meyer2019today}. There is a large body of research that focuses on software developers daily activities and tasks. However, in studying software developers from the same setting and same tasks may be inherently limiting the scope of what a software developer is. Inadvertently, limiting the scope of what a software developer is can also limit who feels they have what it takes to be a software developer.

\subsection{Stereotypes in Computing \& Engineering}

To date, much of the literature around stereotypes in computing have been focused on perceptions on writing code rather than a broader lens on what it means to be developer. 
For instance, Charters et al. \cite{attitudes2014charters} studied the effect of introducing programming to adults through video games as means to change their negative attitudes towards programming. Other researchers have studied the effects of gender, motivation, self-efficacy on people choosing computing as a career and found that stereotypes around computer programming act as a deterrent more than the specific domain (like programming for multimedia) to gender diversity in the world of software developer~\cite{gender2019avilagou, gender2012doube}.
This highlights the impact that stereotypes can have on perceptions of a career at an early stage of interest.
This can also affect on people who are already in their career. Specifically, Liu et al. found that through the use of identity-based hashtag movements engineers could control their personal narrative of what someone in their profession looks like through sharing selfies~\cite{liu2017selfies}. What is also quite interesting to note in the aforementioned work is that this movement started with a software engineer who was told she did not fit the ``cookie-cutter mold'' based on her physical appearance. These works have sparked the conversation of what it means to study stereotypes of marginalized identities. 

\BEGINADDEDX Numerous other stereotypes exist in the software industry. For example Schloegel et. al.~\cite{schloegel2016reducing} discuss how negative stereotypes related to the age of the developers, specifically in certain Asian countries ''can hinder cooperation and team processes, which are of utmost importance in software development''. Such stereotypes risk creating intergenerational tension at work~\cite{levy2016progress}. Stereotypes against minorities in the tech industry, including those based on race and gender, that have created gaps in the diversity of the workforce continue to perpetuate. While recent efforts have seen positive results in reducing gender related stereotypes at workplaces in the software industry, there is little to no change in the ethnic diversities and stereotypes that exist within the software industry~\cite{ethnicdiversity}. In a recent survey on the status of ethnic diversity in the tech industry, 67\% developers report that there is little effort to address the racial inequality within the computing industry ~\cite{ethnicminorities}. \ENDADDED In our work, we expand this intersection of personal and professional stereotypes in the computing community by  studying perceptions around the quality of life, working style, and background of developers who vlog, and the reception of those vlogs.

In order to dismantle stereotypes, one approach that some have taken to are creating videos that can be shared widely. For instance, videos on \yt have been found to be a source where people can post original content where minorities can control the portrayals of their personal experiences~\cite{fung2018feisty}.
Vlogs in particular provide a greater depth to sharing these experiences and also often allow viewers to follow vloggers on a full day of experiences---not just their professional work. This provides viewers with a front row seat at their life. \BEGINADDED This gives vloggers an opportunity to have a discourse with their audience and contradict these stereotypes. \ENDADDED Although vlogs provide this unique perspective, limited research has studied vlogs as a means to challenge stereotypes. \BEGINADDED Researchers have studied the racial and gender diversity among creators and viewers of the vlogs~\cite{meyer2016broadcast, Molyneaux2008ExploringTG}, with the intention to understand their vlogging patterns. Tangentially, Cummings et al. as they studied how one vlogger's channel could provide career advising for African-American students in computing~\cite{cummings2019vlog, cummings2020exploration}. %
Our work expands these discussions within the computing community by not only understanding the various stereotypes that affect the software engineers and developers, \ENDADDED but also understanding the vloggers intentions behind creating this content and how it can impact those who watch the vlogs.

\section{Research Setting: Developers' Vlogs on \yt}
\label{sec:research-setting}

In this paper, we investigate how developers describe a day in their life through vlogs on \yt. \yt can reach a large audience of more than two billion people~\cite{youtubepress} without a gated fee~\cite{burgess2018youtube}. These featured make \yt a supportive and suitable platform to host socio-technical video content. One special type of such socio-technical content are vlogs created by developers. \yt set the trend for vlogs (videos capturing the ``everydayness'' of life) as early as 2005~\cite{duplantier2016authenticity} when Jawed Karim~\cite{jawedkarim}, a software engineer who co-founded \yt, uploaded the first ever video on \yt -- Me at the zoo~\cite{zoo}, a short 18 second vlog capturing his visit to the zoo.

\BEGINADDED We selected to focus on vloggers on YouTube as opposed to other platforms for a couple of reasons. First, ``everybody is already on \yt'' \cite{youtubeLive} with over 2.3 billion users are familiar with the platform~\cite{youtubepress}. Second, \yt has been identified in previous literature as a resource for how to developers find their information~\cite{storey2014revolution}. Finally, \yt is the most common platform for vloggers to the point that recent studies use the term `YouTuber' and vlogger interchangeably~\cite{kidscareer}. While Twitch~\cite{twitch} is \yt's direct competitor, Twitch focuses on broadcasting live sessions. The Twitch community is interested and accustomed to tuning into live sessions and chatting with streamers in real-time. This makes Twitch unsuitable for vlogs, which is rarely in a live-format. Additionally, developers who do live streams on Twitch are switching to stream live on \yt due to Twitch's complex content management system like deleting streams after a certain period, re-uploading live sessions generates a new URL making the live session links posted anywhere on the internet obsolete, etc~\cite{youtubeLive}. \ENDADDED

Prior to 2010, only a few videos existed on \yt that described the activities and skills of developers~\cite{vlogTPPold, vlogFPPoldZune}. It wasn't until the later half of 2010s that social media users frequently started to post vlogs on \yt, a video format which gives the creator a control of the narrative. Further, \yt allows participatory engagement of viewers and creators that leads to multiple interpretations of a shared experience. For instance, the platform provides opportunities for multiple programmers to share their professional work experiences, personal and lifestyle related information, and even provide informal mentorship~\cite{chau2010youtube}. Adopting theory from Lave and Wagner~\cite{lave1991situated}, we understand vlogs as an opportunity to generate a \emph{community of practice} where content creators share a video on experiences of being a developer and viewers engage via the comments to respond to the representation of what was shared. In this setting each video becomes an opportunity for vloggers and viewers to consume, share, and respond to what they consider a developer to be.

\BEGINADDED These vlogs mirror a developer's entire day---from waking up, the food they eat, going to work, what they do at work (like coding, testing, meetings, co-working in teams; Figure~\ref{fig:vlog_at_work1}), breaks, and what they do outside of work (working out, going to movies, spending time with kids and families, or playing games; Figure~\ref{fig:vlog_at_work2}). Each vlog has its own story and emphasizes some of these activities.

\begin{figure}[!tbp]
\centering
\includegraphics[width=\textwidth]{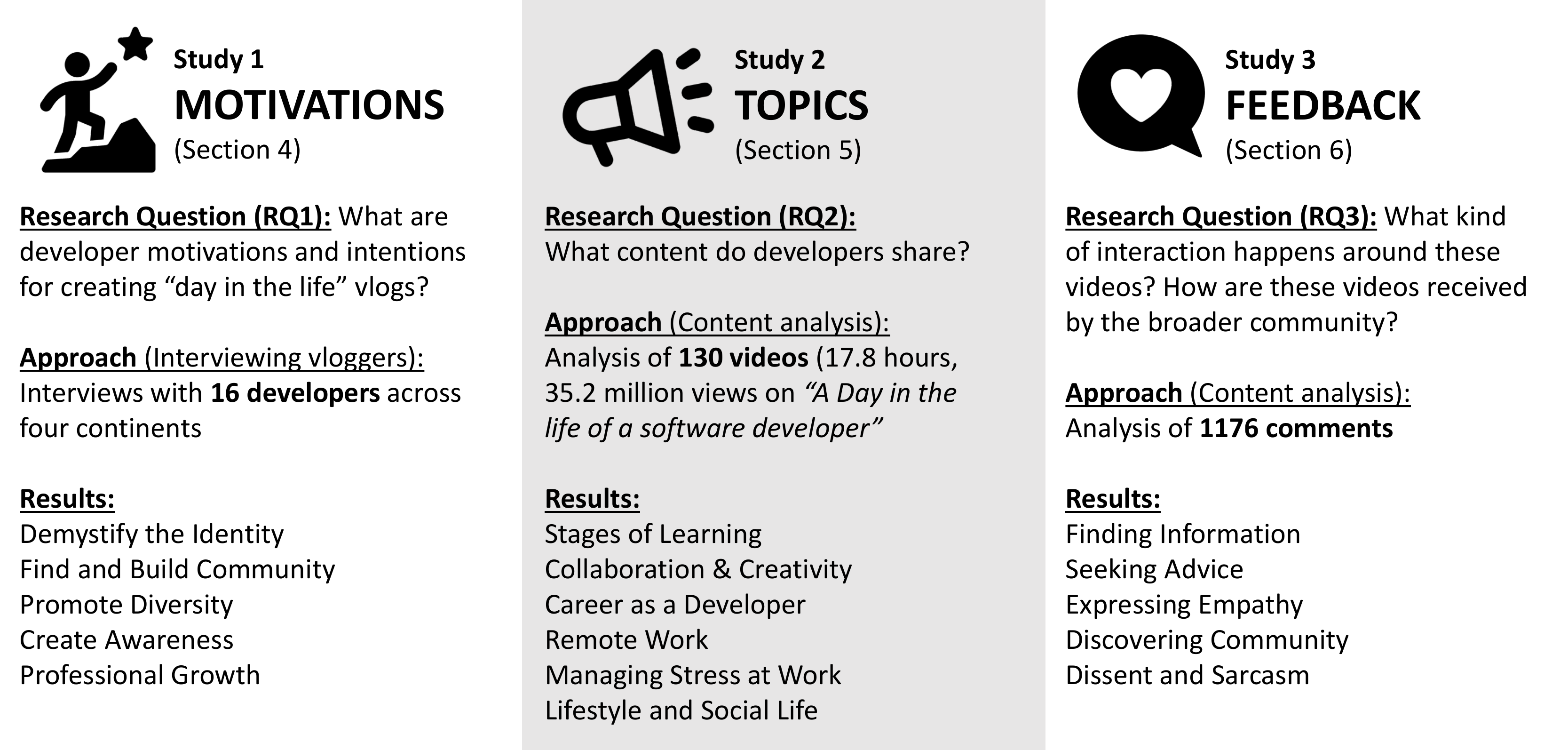}
\caption{\BEGINADDED Three studies to build a converged understanding of why developers vlog and what information is contained in the vlogs. The studies were conducted in parallel.} 
\label{fig:overview}
\end{figure}

In this paper, we studied developers motivations for posting this content (\textbf{RQ1}), how they described a day in their life (\textbf{RQ2}), and understand the viewers' reception of the vlogs through comments (\textbf{RQ3}). \ENDADDED We conducted \textbf{three studies in parallel} to build a converged understanding of \emph{why} and \emph{what} information is contained in the vlogs (see Figure~\ref{fig:overview} for an overview). In our studies, a \emph{vlog} refers to one video from an individual developer on \yt. \BEGINADDED This research was reviewed and approved by a federally registered Institutional Review Board (IRB) \emph{[IRB name and number redacted for review to preserve anonymity]}. An IRB is an entity that ensures that research studies are executed in an ethical and compliant manner.
The supplemental materials package can be found in the online submission portal, which contains the full interview script, the list of vlogs used for this research, and the qualitative codebooks used to analyse the videos and comments. \ENDADDED

\begin{tcolorbox}[   %
        breakable,
        left=1pt,
        right=1pt,
        top=1pt,
        bottom=1pt,
        colback=gray!13,
        colframe=black,
        width=\dimexpr\columnwidth\relax, 
        enlarge left by=0mm,
        boxsep=5pt
        ]
\begin{description}\BEGINADDED 
\setlength{\itemsep}{0pt}
\setlength{\parskip}{0pt}
    \item[RQ1] \RQA
    \item[RQ2] \RQB
    \item[RQ3] \RQC
\end{description}
\end{tcolorbox}

\smallskip
\BEGINADDED In the following sections, we present three studies to answer each research:  Study 1 investigating the motivations for creating these specific vlogs  (Section~\ref{sec:study-interviews}), Study 2 understanding the content in these vlogs (Section~\ref{sec:study-vloganalysis}), and Study 3 understanding the influence of these vlogs (Section~\ref{sec:study-comments}). We use \textbf{Findings}, green boxes annotated Finding~\ref{takeaway1}--\ref{takeaway23}, to highlight the key findings throughout these three sections (Section~\ref{sec:study-interviews}--\ref{sec:study-comments}). We present a Discussion (Section~\ref{sec:discussion}) of our findings across the three studies and refer to the Findings boxes when doing so.
Finally, we discuss the Limitations (Section~\ref{sec:limitations}) of our research, and present our Conclusion (Section~\ref{sec:conclusion}). \ENDADDED

\BEGINADDED
\section{ Study 1: \RQA (RQ1) }
\label{sec:study-interviews}
\ENDADDED

\BEGINADDED In the previous section (Section~\ref{sec:research-setting}), we described that developers share a day in their life through vlogs on YouTube. In the first study, we investigated through interviews the motivations that software developers have to create these vlogs.\ENDADDED %

\subsection{Approach} \BEGINADDED To understand why developers vlog about their everyday life, we interviewed 16 developers who upload vlogs on \yt. We identified the vloggers from a list of of developer vlogs that we collected for another research question. Each interview participant consented to participate in the study online before beginning the interview. 

\paragraph{Recruitment.} We used two different methods of recruitment. We sent emails (or messages) containing a sign-up link to the email addresses or social media profiles listed on the \yt channels of the developer we identified in Section~\ref{sec:1a}. We then posted the sign-up link on Twitter, openly inviting developers who vlog. The sign-up link contained demographic questions like age, gender, years of work experience, and a link to reserve a time slot for the online interview. Participants were compensated with a gift card worth 25 USD.

\paragraph{Participants.} We had 23 developers signed up for the interviews, and 16 of them completed the interviews. Table~\ref{tbl:participants} shows their location, age and gender, years they have been programming and vlogging, and their current job titles. Each participant is referred to throughout the paper as P\#.
7 out of 16 participants identified themselves as women, 9 as men. Our participants joined from North America, Europe, Asia and Middle-East. All participants  held some position equivalent to (or higher then) software developer/engineer at some point. \participant{P12} is in the middle of transitioning jobs and held no position at the time of interviewing (marked as NA). Three other participants have quit their full-time jobs and become full-time creators on \yt due to circumstances surrounding family \participant{P04}, or personal choice of career. 

\paragraph{Protocol.} The semi-structured interviews lasted between 30-45 minutes. We asked the participants about their motivation for starting a \yt channel and why they continued to post content regularly. We also asked them about what factors they consider when creating the vlogs and the effect they perceive these videos have on the general audience. The interview guide is included in the supplemental materials package submitted in the online portal.

\paragraph{Analysis.} We transcribed the interviews and assigned descriptive codes categorize the different motivations and factors, similar to the process of analyzing the videos. Two authors then performed axial coding by iteratively grouping and re-grouping similar descriptive codes, and redefining the groups into high-level themes. We reached saturation in the interview analysis for the high-level themes after 16 interviews, i.e., no new themes appeared in the final interviews.  %
\ENDADDED 

\begin{table*}[!t]
\renewcommand{\participant}[1]{P\lookupGet{#1}\xspace}

\caption{\BEGINADDED Interview Participants. The \textsc{Age}, the experience programming (\textsc{Exp$_{\textit{prog}}$}), and the experience vlogging (\textsc{Exp$_{\textit{vlog}}$}) is in years. \label{tbl:participants} \ENDADDED}
\resizebox{\textwidth}{!}{%
\begin{tabular}{@{}clccccll@{}}

\toprule
 \textsc{ID} &
 \textsc{Location} &
 \textsc{Age}&
 \textsc{Gender} & 
 \textsc{Exp$_{\textit{prog}}$} &
 \textsc{Exp$_{\textit{vlog}}$}&
 \textsc{Job Type}&
 \textsc{Job Title}\\

\midrule 
\participant{P02} & Japan & 32 & M & 7 & 1 & Remote & Software Engineer\\ 
\participant{P04} & USA & 33 & W & 8 & 1 & Freelance & Creator\\
\participant{P05} & India & 23 & M & 5 & 2 & Remote & Software Engineer\\
\participant{P06} & UK & 33 & M & 20 & 2 & In-office & Software Development Engineer\\
\participant{P08} & Spain & 25 & M & 15 & 2 & In-office & Software Engineer\\
\participant{P11} & Philippines & 24 & W & 4 & 2 & In-office & Front-End Web Developer\\
\participant{P12} & USA & 24 & M & 7 & 2 & NA & NA\\
\participant{P13} & India & 23 & M & 7 & 1 & In-office & Senior Developer\\
\participant{P14} & USA & 23 & M & 4 & 0 & In-office & Full Stack Software Engineer\\
\participant{P16} & Iraq & 26 & W & 7 & 0 & In-office & Developer\\
\participant{P17} & USA & 25 & M & 6 & 2 & Freelance & Developer\\
\participant{P18} & USA & 27 & M & 3 & 2 & In-office & Software Engineer\\
\participant{P20} & USA & 24 & W & 4 & 1 & Freelance & Creator\\
\participant{P21} & USA & 27 & W & 10 & 3 & Freelance & Creator\\
\participant{P22} & USA & 28 & W & 4 & 1 & In-office & Software Engineer\\
\participant{P23} & Germany & 26 & W & 6 & 0 & In-office & Engineer\\
\bottomrule
\end{tabular}
}%

\renewcommand{\participant}[1]{\textcolor{darkgreen}{P\lookupGet{#1}}\xspace}
\end{table*}

\subsection{Results}
\label{sec:rq1_res}

\BEGINADDED
From our interview analysis, we found that developers were motivated to vlog to disseminate various misconceptions about ``who are developers'' within the different groups of people and posted the vlogs on \yt as it reaches a diverse audience. Through the vlogs, developers wanted to unveil what software developers do in reality, reach out and build a community, promote a diverse set of experiences within that community, and create awareness of the variety of opportunities in this field of work. We discuss each motivation here:
\ENDADDED

\subsubsection{Demystifying the identity and life of the developers:}
\label{subsec:why:identity}
\label{legacy:5.1}
\label{legacy:5.2.1}

Many developers who vlog want to demystify what it means to be a developer; the image people have about developers is \inlinequote{very intimidating} \citeParticipant{V11} and developers are not seen \inlinequote{in a human light} \citeParticipant{V22} and instead as \inlinequote{mythical creatures} \citeParticipant{V16} who \inlinequote{glamorize the coding and life and big tech companies} \citeParticipant{V12}. They use their videos as a tool to counter these stereotypes.

\participant{V20} wants to show her viewers that developers are just like other people and sends the message that \inlinequote{I am very far from perfect, and that there are a lot of bumps I've had in my journey and that I still experience.} In her videos, \participant{V21} uses the theme that it is possible to \inlinequote{be a developer and be successful in your career and have a fulfilling, meaningful life in other ways.}

\takeawayboxx{takeaway1}{\BEGINADDED Through vlogs, developers relay that they are human who lead a rather simple life beyond their monitors, much like other professions. Coding is a skill they acquire through knowledge and practice over the years.\ENDADDED}

With the abundance of videos about developer life on \yt, there are some videos \inlinequote{that are inaccurate or just plain out seeking attention, which is not necessarily what actually happens} \citeParticipant{V06}. This motivates many developers to create videos that are more ``realistic''. For example, \participant{V06} talks about how \inlinequote{there are many videos out there that touch on the subject of software development and programming, and [yet] they talked about different stuff and they are often very poorly made.} \BEGINADDED Such videos portrayed developers as super-beings who can plan out and use every minute of their lives to build amazing tools and softwares, whereas, in reality developers are \inlinequote{living, breathing human people who are dynamic and have their own lives and have interests outside of coding} as \participant{V21} shows in her videos.

A common measure to overcome this misconception was to show what developers actually do as part of their work. \ENDADDED  Through her videos, \participant{V22} wants to challenge the expectation that developers code all day. \inlinequote{I try and show that I do other things besides coding. I'm not just somebody who sits like a robot, who just sits and codes all day.}
\participant{V12} is concerned about the glorification of coding and is \inlinequote{trying to point out that [developers] are not having fun all the time.} \participant{V11} wants people to know that developers can \inlinequote{also enjoy and learn.}

To emphasize their life outside of work, developer who vlog talked about their hobbies and interests. \participant{V02} has an \inlinequote{interest in weight training and fitness} and enjoys reading content online, \inlinequote{it doesn't really involve just computing or development. It can be anything, for example, history.}
Many come from artistic background: \participant{V06} recalls \inlinequote{I majored in Fine Arts with traditional painting as my main subject}, \participant{V04} is a creative person and enjoys activities like \inlinequote{creating a t-shirt or birthday card}, \participant{V11} pursues \inlinequote{photography as a hobby}, and \participant{V17} enjoys \inlinequote{ballroom and cultural dancing.} \participant{V23} enjoy \inlinequote{pretty basic stuff, nothing super crazy. Just normal usual stuffs that all the people do.}

\takeawayboxx{takeaway2}{\BEGINADDED Developers emphasize through the vlogs that they don't code all day, and have interests and hobbies beyond coding (or work in general)---from arts to sports.\ENDADDED}

\BEGINADDED The general public find developers intimidating and often don't see them in human light. To counter this perception, developer show other ``human'' aspects of their life beyond coding and share their diverse backgrounds and hobbies.\ENDADDED

\subsubsection{Finding and building a community}
\label{subsec:why:community}
\label{legacy:5.2}
\label{legacy:5.2.2}

\yt provides a platform for developers to connect to other people.
For example, \participant{V06} used to be part of an online community around a gaming project, \inlinequote{I missed having a community that is gathered around something that we built together. So I decided to pick up videos as a new form of expression.} \participant{V11} uses his \yt channel to \inlinequote{reach out to other people, share what I'm doing in my life or what are the experiences that I've been in my career.} \participant{V17} uses \yt as a platform to connect to peers.  For \participant{V20} it is the sense of belonging, \inlinequote{to hear other people's stories and be inspired by them. That's definitely what feels the best part of this entire experience, is just interacting with my community.}
\BEGINADDED For example, \citeParticipant{V17} met another developer through his channel, and after connecting through Instagram, they collaborated together to create an online course and also posted another vlog of them working together on \yt. \ENDADDED

\takeawayboxx{takeaway3}{\BEGINADDED Developers who vlog build a community of their own, learn and share experiences with this community and interact with each other. They also collaborate to create contents for \yt or other ventures like co-creating courses, freelance groups, and applications.\ENDADDED}

Viewers also create a community for the developers, and also are the motivations behind some of the content the developers produce. Viewers reach out to the developers with questions like ``What do I learn the most?''. Other viewers reach out to the developers to thank them \inlinequote{for helping me figure out this and understand that it's not only me who is feeling like an imposter} \citeParticipant{V04}, or inspiring the viewers \citeParticipant{V20}. With time, developers continue to vlog to support their community: \inlinequote{I don't want to be the reason you got into tech and then not be able to help you navigate} \citeParticipant{V22}. 

These viewers include students from various fields. \participant{V18} uses his channel to share the experience of being an international student who became a developer, and most of his viewers reach out with educational questions and are \inlinequote{people who are in Korea preparing or wanting to come to America and are curious about the SAT process, APE credits, TOEFL tests, how to find the colleges, and what major they should have.} 
Viewers also include younger who are looking to make a career switch into technology. As \participant{V05}, explains some viewer \inlinequote{have a choice where, they feel that what they're doing isn't worth doing and won't lead them anywhere. They are still able to branch out to technology.}
Occasionally, viewers are \inlinequote{people who are totally outside of this world and they are just curious what's happening inside the technology world.} \citeParticipant{V06}

\takeawayboxx{takeaway4}{\BEGINADDED Vlogs are a source for viewers to connect with other viewers as well as the developers posting vlogs. Viewers interact directly with developers through comments or other social media platforms, and connect with other viewers going through shared experiences for moral support. \ENDADDED}

\BEGINADDED Vlogs can be the foundation for different communities --- developers who upload vlogs find like minded developers to collaborate and connect with, and viewers also find communities within developers and other viewers who interact with the videos with similar goals.\ENDADDED

\subsubsection{Promoting diversity within the developer community}
\label{subsec:why:diversity}
\label{legacy:5.3}
\label{legacy:5.2.3}

Many developers who vlog are driven by the motivation to create a more inclusive community. Vlogs can \inlinequote{help people find their way to the industry, especially like people who are minorities in tech} \citeParticipant{V21} and  \inlinequote{because sometimes tech can be overwhelming} \citeParticipant{V04}.

\participant{V21} started vlogging because as \inlinequote{a woman in tech, a constant minority in every scene in tech}, she could relate to the cause on a personal level. \participant{V04} is a mother and has seen that many women give up their career as a developers to focus on their responsibility as a mother, she wants to encourage women in parenthood through her vlogs. \inlinequote{I wanted to say, look, this is possible, you can be a parent and you can work as a Developer or any other position if you want.} \participant{V04} also shared her experience volunteering at ``Women who code'' because she \inlinequote{found that if I see somebody else's experience, I don't feel out of the loop.} \participant{V22} says her motivation is to get ``free resources out there, to help minorities specifically get into development.'' 

\takeawayboxx{takeaway5}{\BEGINADDED Women constantly feel like a minority in coding communities. These women developers encourage more women to pursue computing and include links to helpful resources and support systems in their vlogs. \ENDADDED}

The region of a developer often plays a role in promoting diversity and \yt creates a platform for others to relate to the developers who vlog. 
\participant{V11} focuses on promoting diversity within Philippines as she says \inlinequote{most of the developers here in the Philippines are men. I'm also trying to tell the viewers that girls can also code, girl power. I also want to represent women who code, I want to encourage them to just continue doing this until they get the job.} 
\participant{V16} from the Middle-East wants to \inlinequote{encourage other girls and other people like me. [\dots]. I know the environment is not very female-friendly for software engineers, but we can do it!} Her motivation comes from being able to encourage girls to pursue this career. \inlinequote{If one girl gets encouraged to learn something from my videos or even just learn about CS, then the job is done.}

Barriers to diversity sometimes include factors such as misconceptions and social norms. 
\participant{V16} explains \inlinequote{There [are] misconceptions about a software engineer. There are misconceptions about being a Muslim. There are misconceptions about being from the Middle East.
[\dots] I can put it in something that is somewhat humorous, somewhat informative, I hope that the exposure can be informative}
\participant{V20} wants to challenge the norms about a reputable career in the Indian society and \inlinequote{I used to be pre-med and I used to work as an engineer, but I left all of those things to have a YouTube channel. Which is very atypical and not according to the standard expectation of, I guess, an Indian parent. But [I want to] show that that's okay, and share my path along the way.}

\takeawayboxx{takeaway6}{\BEGINADDED Developers also use their vlogs to talk about regional and religion based misconceptions about computing careers across the world. They share their personal struggles of overcoming these barriers, and encourage the viewers.\ENDADDED}

\BEGINADDED To summarize, developers use vlogs to advocate for the minority in coding communities. Women developers who vlog feel the need to increase diversity within their communities \citeParticipant{V22}, and some of them take to vlogs to inform and relay personal stories of overcoming the barriers of diversity. Through the vlogs, developers inform minority communities about resources and support systems.\ENDADDED

\subsubsection{Creating awareness for software development as a career opportunity}
\label{subsec:why:influencing}
\label{legacy:5.4}
\label{legacy:5.2.4}

Through their vlogs, several developers want to create awareness about software development as a career and how to get into programming. 
\participant{V05} uses his \yt channel to \inlinequote{spread the message that anyone can basically learn to code and get into tech} and creates videos and vlogs about \inlinequote{the technologies and how to learn these things and how they can help.}  
\participant{V14} shares \inlinequote{experiences as someone who left college and started working in the software engineering industry} because it can eventually \inlinequote{help someone not make the same mistakes that I did and will be making in the future.}
\participant{V06} creates vlogs to make \inlinequote{people interested in our world about software development and being a programmer and [how it is like] joining a company of this size.} 

\participant{V18} wants \inlinequote{to share [his] experience being an international student and also someone in tech in the US} in order to make people aware about opportunities abroad because \inlinequote{some of my friends back in Korea don't really know about [these opportunities].}
His motivation also comes from his experience of not having information about how to pursue programming and development as a career. Through his videos he wants \inlinequote{people who are interested in becoming a software engineer in the future to get a sense of what it's like, and what to expect, and maybe (if they're interested) how to prepare for that career path as well.} 

\takeawayboxx{takeaway7}{\BEGINADDED Coding is not only for the chosen ones. Developer use their vlogs to relay that coding is a skill that can be acquired through practice and there are many non-conventional opportunities to learn programming and pursue a career in computing.\ENDADDED}

\BEGINADDED \participant{V02} started vlogging to create awareness about work culture and career opportunities in Japan. \inlinequote{I was looking for a lot of vlogs on remote development, particularly in Japan. What I found was there was not really a lot of content or ideas about that. I wanted to make something unique on YouTube, and I thought I could share to the world, especially in Japan, where a lot of the work culture is quite traditional, what the benefits could be where people can work remotely.} \BEGINADDED

\participant{V05} is a remote developer from India who talks about the \inlinequote{misconceptions and misunderstanding about what technology is like}. Some Indian IT companies \inlinequote{would take engineers and they would get them to work and not really pay [well]}. This creates a barrier for people to pursue software development as a career. However, \participant{V05} \inlinequote{realized the opportunities that are currently available} and uses his \yt platform to inform people about the various career opportunities in India. 

\takeawayboxx{takeaway8}{\BEGINADDED Career opportunities and working conditions across specific countries are unique, and developers share their experiences and knowledge about norms within these industries through their vlogs.\ENDADDED}

\BEGINADDED
Through the vlogs, developers talk about the different aspects of career in software development and how to get started. Some developers emphasize regional opportunities and speak out on different misconceptions about career as a developer.
\ENDADDED

\subsubsection{Professional Growth}
\label{legacy:5.5}
\label{legacy:5.2.5}

Vlogs often lead to career opportunities for developers in the long run. \participant{V06} uses his  channel share positive experience about the companies he works for and makes sure that he doesn't use his platform in aggressive ways. \inlinequote{By doing so and behaving in a professional manner, you're actually growing your career and it's pushing your career forward. That's actually a proof, because I was offered a job in Cisco, out of the blue. Probably hugely influenced by these type of vlogs} \citeParticipant{V06}. Developers also find \yt as a platform that creates \inlinequote{new ways of connecting with different people and spreading personal social brands and online brand} and for \participant{V08} this made \yt a \inlinequote{perfect window for making connections}. 

Developers who vlog also gain skills, for example, \participant{V08} gained  \inlinequote{communication skills, like marketing skills, public speaking skills} as well as other problem solving skills like \inlinequote{organizing the thoughts and see what are the consequent or the sequential steps that are needed.} Creating vlogs helps building confidence in skills as a programmer, as \participant{V08} explains, \inlinequote{{}when mentoring junior people in my company, and I need to walk through different pair programming sessions or different problems, the fact that I'm very used to anonymous people watch me code makes me feel more natural.} \participant{V21} says vlogging has helped her become a better learner by \inlinequote{%
being able to learn how to do new things by reading documentation and researching.} Whereas for other developers, vlogs are a way to create a personal diary. \inlinequote{I like being able to go back and seeing what I've done} and vlogs give the \inlinequote{ability to go back and see the things that I've done. I like having it they're documented} \citeParticipant{V14}. 

\BEGINADDED
From these five motivations we see that developers are motivated to vlog to help people at different stages of their life. It starts with clarifying identity of developers among the population who hold certain misconceptions against developers, providing a starting point for the minority communities to find resources and support, and explicitly relaying personal stories to promote further diversity within the computing community. Finally, developers are also motivated to provide information and awareness through their vlogs to people who are aiming to get into the computing career and those looking at computing careers in the long run.
\ENDADDED

\mybox{\faArrowCircleRight~\textbf{Summary:} \BEGINADDED Developer motivations to vlog are breaking misconceptions against developers (Section~\ref{legacy:5.2.1}) and helping minority communities around the world overcome barriers (\ref{legacy:5.2.2}). Developers are also motivated to promote overall diversity within the computing community (\ref{legacy:5.2.3}). Finally, developers use the vlogs to provide useful resources and information to people trying to get into the computing career (\ref{legacy:5.2.4}) and those who are looking into a long term career in this field (\ref{legacy:5.2.5}).\ENDADDED}

\BEGINADDED
\section{ Study 2: \RQB (RQ2)}
\label{sec:study-vloganalysis}

From Study 1, we find that vlogs presented an opportunity to demystify the full life of software developers and bring awareness to the diversity of the role and who can pursue this career. In our second study we investigate how software developers present this awareness.
To answer this research question, we explored what types of information developers shared as part of their vlogs on \yt.

\subsection{Approach}%
\label{sec:1a}

\paragraph{Data Collection.} We iteratively searched \yt to identify 163 vlogs using a stratified sampling method to represent countries/regions with a strong developer presence. We identified the initial set of videos through \yt search using the keywords ``[developer life'', ``day in life'', ``day in life + software engineer'']. North America and Europe was well represented in the initial sample. To extend the representation to Asia, we conducted a focused search with the modified keywords [``developer life + korea'', ``developer life + japan'', ``developer life + india'']. This search also lead us to more videos from other Asian countries like Singapore and two videos from China as well. Further, we identified 8 more videos through snowball through (1) \yt recommendations that revealed vlogs that were not part of the initial sample, and (2) videos and vloggers mentioned within vlogs within our sample.

Our final sample included 130 videos from the 163 videos. We excluded 33 videos after initial analysis as 6 videos were not in English, 14 videos appeared in the selected list more than once, 7 videos were made by students or interns as revealed within the video or their channel description, and 6 other videos were not vlogs but other in other forms like a recording of conference talk about a developer's life. 103 videos were uploaded by men, 24 videos by women (1 featured a couple, and for 2 other videos we couldn't determine the gender).

\paragraph{Analysis.} To analyze the diversity within the videos, we manually identified and compiled standard \yt metadata from each videos including the date of upload, length of the video, location, subscriber count, view count, likes and dislikes count, number of comments and the first 10 comments from each video (sorted on \yt by most engagements on the comment e.g., likes form other viewers, replies on the comments). All 130 videos were uploaded within the last 4 years (between January, 2016 and May, 2020), and follow the standard length of \yt videos~\cite{clement2019youtube} with an average length 8.23 minutes. The 130 videos are from 113 developers who vlog from 21 countries across the four continents of North and South America, Europe, and Asia. Each vlog is referred to throughout the paper as [V\#]. %

\yt as a platform has the advantage of reaching a wider audience beyond just the subscribers. To identify the reach of the vlogs we analysed the metadata associated with viewer engagement. It is difficult to identify and measure this population, since \yt doesn't provide identity of the viewers. However, we estimate the breadth of the population through the subscriber count for each channel, and the view counts on each video (assuming each viewer watches a vlog one or two times only). The channels of the 113 developers had an average of 143,993 subscribers (min. = 3, max. = 2.41 million, s.d.= 393K). On average, each vlog had 270,800 views (min. = 83, max. = 7.131 million, s.d.=898K). Thus, the information within these vlogs reach a large audience.

To analyze the what information the vlogs contain, the first author transcribed the videos and assigned descriptive codes (labels/short phrases)~\cite{Saldana2009} to the various topics covered by vloggers as well as the activities they show as part of their everyday life. All three authors then collaboratively reorganized these codes and performed selective coding~\cite{Saldana2009} by grouping related topics into stand-alone thematic concepts. The authors met multiple times over the next few weeks merging, splitting, and reorganizing the topics to identify the themes that describe ``what is a developer's life about'' as expressed in the vlogs.

\ENDADDED
\subsection{Results}%
\label{sec:Topics}

From our conceptual analysis of vlogs, we find that developers discussed experiences at the intersection of their professional and personal life. We identified six categories including their learning journey, collaboration \& creativity of software development, their career trajectory as software engineer, non-traditional work settings, stressful challenges faced at work, and finally their lifestyle. The topics covered below are ordered from highest to lowest prevalence in vlogs. %

\subsubsection{Stages of Learning}
\label{legacy:6.1}

\paragraph{Learning different programming languages.} Vlogs also shared guidance and tips on how to enter the world of programming through determining which programming language. This often started with what programming languages are great first languages to learn \citeVideo{M1_23}, \citeVideo{M1_39}, \citeVideo{M1_47}, \citeVideo{M2_125}. Vlogs also described that often times the programming language you use is based on the task and provide examples from their professional experience.
For instance, one vlog lays out how often they determine the programming language to use: ``in terms of what coding languages I use at work it really does depend on the task. My team works across a ton of business units so we're always having to switch. I mainly use JavaScript for our node-based applications, and then I use a ton of Scala as well'' \citeVideo{M2_125}. Another vlog points out that languages like Java, ``help to master fundamental concept of a programming languages such as data types, object oriented control statements, data structures and algorithms, reading network protocols and the graphic debarment'' \citeVideo{M1_39}. Vlogs also go on to describe the different roles within development---front-end which requires ``HTML, CSS, along with some JavaScript'' \citeVideo{M1_23} and back-end where one needs to use ``C\#, SQL and different databases like MongoDB'' \citeVideo{M1_23}. This particular vlog also describes the benefits of being a full-stack developer, skilled in both front and back end, as ``having that jack-of-all-trades is great because you could work at a smaller company where you can be very very effective and you can bring a lot of value to the table'' \citeVideo{M1_23}.

\paragraph{Self-learning.} 
The changing perception of software developers' identity and work is further emphasized when developers share their journey of self-learning as part of their career transition in their vlogs (\citeVideo{MS_19}, \citeVideo{M1_47}, \citeVideo{M1_82}). %
Developers shared their journey of learning programming through online platforms like CodeAcademy, TreeHouse\cite{treehouse}, W3schools.com, Coursera and discuss their good and bad experiences. While one developer shared getting a degree from Teamtreehouse.com lead to him being hired \citeVideo{M1_03}, another developer says he tried ``CodeAcademy for a little while and got bored of it'' due to the overwhelming amount of content \citeVideo{M1_47}. Developers provide various tips on effectively navigating the contents to successfully learn the techniques \citeVideo{MS_19}, and how to polish those skills by practical applications and bootcamps \citeVideo{M1_33}, \citeVideo{M1_47}.

\takeawayboxx{takeaway10}{\BEGINADDED Vlogs discuss different programming languages to start the programming journey, discuss various areas and roles within software development, and their experiences and knowledge of alternate pathways to pursue development career.\ENDADDED}

\subsubsection{Collaboration \& Creativity}
\label{legacy:6.2}

A developer in one of the vlogs argued that one reason software development is perceived to be more difficult than it is in real life is because ``people have a misconception that all you do is sit in front of a screen and stare at a text editor (for code) all day'' \citeVideo{M2_99}. Whereas in reality a developer need to know more than just how to code. They then clarify: ``I would actually say that not a lot of my time is in front of an editor! A lot of it is talking to people, and figuring out what we should actually be doing, coordinating meetings, writing minutes, \ldots, [even] sometimes just sitting in my chair thinking about what to do!!'' \citeVideo{M2_99}. %

Another developer described how being a developer gives them a chance to be innovative ``[this] gives you a chance to be technical and creative. You get the technical aspect of doing the programming and solving problems. But you also get to have a bit of a creative flair and express yourself a little bit'' \citeVideo{M1_34}. Developers were described as having to be ``very comfortable'' with the technologies they need to use \citeVideo{M1_48} while also being able to follow ``logic and problem-solving'' \citeVideo{M1_40}. In addition, several videos described how it important it was for developers to have adequate communication skills since, ``often you can get a lot more done by having a quick conversation with the rest of your team'' \citeVideo{M1_40}. 
In short, vlogs positioned software development as more than staring at the editor---it is the amalgamation of technical creativity and communication skills that make a successful developer.

\takeawayboxx{takeaway12}{\BEGINADDED Development is more than coding, as developers spend a lot of time talking to co-workers and designing solutions to the problem. Vlogs discuss communication and creativity are an important part of developer's work.\ENDADDED}

\subsubsection{Career as a Software Developer}
\label{legacy:6.3}

\paragraph{Who is fit to be a developer?} In the vlogs, we find that developers break down the public perception of required skills to pursue a career in software development. We identified 28 videos that highlighted non-traditional pathways and skills into a development career. In one of the vlogs, a developer who made a late career change to software development emphasized that people do not have to be an expert in mathematics to enter software development: ``I'm just a guy. I'm not a genius, I don't know trigonometry or calculus or an advanced math. I went to school originally for creative writing. I want you to know that just because you're not X or Y, you can still do this'' \citeVideo{M1_47}. In another video, a developer who called themselves ``just a film school graduate living in the middle of nowhere'', shared his experience of transitioning from an art background to the world of development \citeVideo{M1_82}.\BEGINADDED A developer shared in her vlog that after she learned some of the concepts and terminology on her own, ``she researched career paths to see if this is something that I would be interested in'' \citeVideo{MS_19} and eventually joined an online technology school for programming to take various courses.\ENDADDED

\paragraph{Career trajectory and expectations.} Developers in the vlogs also discussed their career trajectories. A developer points out ``the norm when you work for a new company as a front-end developer or engineer is, you don't work on tough things right away but you kind of get put into [easier tasks] to get more comfortable and then when the time comes it goes crazy!'' \citeVideo{M1_03}. Along the journey of development ``every new requirement comes with a new challenge. One requirement can come with a lot of flexibility in designs, while another can come with [learning] new technology. While working on different challenges you grow yourself as a professional, as a developer'' \citeVideo{M1_48} says a senior developer. Another senior developer talks about the decisions developers take in their career, ``as you grow you also start to manage some people and mentor some people and look after you know small to mid-size teams \ldots{} this crossroads is pretty much asking you do you like managing people more than engineering? or do you like engineering more than dealing with people?'' \citeVideo{M2_123}. At times, these transitions are accompanied by a change in company. A developer, who's transition from developer to senior developer required changing companies, discuss the various factors that are involved in those decisions like ``scope of services and components [he will be working with], interest in leadership positions, and the timing of project completion'' \citeVideo{M1_78}. Finally, we found that vlogs shared personal stories about being a developer which contribute to breaking down the perceptions about how people grow in software development positions, thus turning their path into ``a career as opposed to just a job'' \citeVideo{M1_40}.

\takeawayboxx{takeaway14}{\BEGINADDED Developers discuss their journey from early years, as they grow constantly learning new technologies with every new projects into their senior years. As senior developers, they discuss how to handle decisions like taking on managerial roles or transitioning between companies.\ENDADDED}

\paragraph{Freelancing.} Developers in vlogs discussed the the advantages and disadvantages of freelancing, a common style of work for software developers. An advantage mentioned by freelance developers was the a great deal of flexibility their position provided. It gave them freedom to ``create your own schedule'' \citeVideo{M1_26}, and control their own time \citeVideo{M1_02}. This impacts the productivity at work; a developer shares it gives him the freedom to ``go for a run now clear my head get a mental break instead of doing it later'' before getting into mentally taxing refactoring work \citeVideo{M1_26}, \citeVideo{M1_11}, or take ``the kids to the Botanical Gardens'' \citeVideo{M1_22}. The flexible time schedule accommodates spending more time with the family \citeVideo{M1_33}, and even some time for themselves \citeVideo{M1_11}. Freelancing work also often allowed working remotely from a variety of locations. A prominent challenge for freelancers was having to switch context of code bases as was described in one vlog: ``switch code bases quite often and there's always a ramp up time when you're learning a new code base trying to identify the issues that you have.'' In this vlog she gives an example of how sometimes the clients use ``different versions of Ruby and rails,'' which requires additional effort and time to set up on her machine \citeVideo{M1_22}. Another vlog included a freelancer describing the stress that can be associated with short-lived nature of projects since contracts are ``built on lease, not working means not getting paid'' \citeVideo{M1_22}. These challenges span beyond physical work setting and are relevant to general work productivity. To support productivity, developers  also recommended many tips and best practices in their vlogs. 

\takeawayboxx{takeaway15}{\BEGINADDED Freelancing is common style of work among developers today, and developers discuss their experiences of working as a freelancer when balancing family, multiple jobs, making time for other interests etc.\ENDADDED}

\subsubsection{Remote Work} 
\label{legacy:6.4}

Developers discussed working remotely along with associated perks in the vlogs. Remote work allowed the developers to work from a variety of locations. For some developers, it is from the comfort of home at ``your own'' work desk and equipment \citeVideo{M2_94}. This saves time lost in transit to work \citeVideo{M1_26}. For others, remote and freelance work allows them to ``just take the laptop and pretty much go to different cities and just work from there'' \citeVideo{M1_17}; like working from Florida instead of Chicago in the winter conditions \citeVideo{M1_26} or working from Berlin for a change of scenery \citeVideo{M1_20}. 
When developers discussed working remotely they often discussed how ``you can get distracted easily'' when working from home \citeVideo{M1_68} and thus many often considered renting ``co-working spaces'' \citeVideo{M1_22} or finding  coffee-shops \citeVideo{M1_26} and public lounge spaces \citeVideo{M1_30} to work from to stay productive.

\paragraph{Quarantine \& COVID-19.} 
As we conducted this video analysis during the summer of 2020, our sample of vlogs included videos in which developers shared how they work during a global pandemic and the additional challenges to productivity they face in a time of crisis. In these vlogs, there were many developers who were already working remotely and several who were transitioning to remote work and described their challenges. Some of these challenges included non-ergonomic work settings \citeVideo{M2_125} and challenges faced to their mental well-being \citeVideo{M1_75}.
Specifically for freelance workers, the pandemic introduced projects being cancelled since some organizations ``don't have enough money to pay all the engineers and the remote engineers'' \citeVideo{M2_127}. Videos also included approaches on how teams were identifying ways to overcome challenges such as remaining socially-connected with their team via ``remote lunches'' \citeVideo{M1_75} over a video call. These vlogs in some ways serve as a time capsule of how software developers worked during that complex stage of remote work.

\subsubsection{Managing Stress at Work}
\label{legacy:6.5}

\paragraph{Stress and Exhaustion.} Not very often, but developers open up about the stressful nature of their work and exhaustion in their vlogs. One developer describes the daily life of a developer involves ``a lot of stress about why code is not working today'' \citeVideo{M2_100}. Another developer captures a similar scene of confused expression of himself with his head in his hands while trying to make sense of the code \citeVideo{MS_16}.  Although reviewing code filled with errors was quite frustrating, developers also presented ways to cope such as time boxing how long they will try to understand a code snippet before taking a break: ``I use the Pomodoro Technique which is 25 minutes on five minutes off.'' \citeVideo{M1_22}.

Likewise, long stressful working hours can lead to exhaustion at the end of the day, as a developer states ``now I had a really kind of busy morning, I'm kind of mentally drained'' \citeVideo{M1_26}.  %
To manage stress developers also capture how they worked on personal projects after work like making an iOS Apps called Dinner Roulette using Swift at home after work hours \citeVideo{M1_19}. Others mentioned that to get their mind off work they do things that are not at all related to software development. One vlog highlighted how one software developer coded during the day, but was a pastry chef by night \citeVideo{M1_27}.

\takeawayboxx{takeaway17}{ \BEGINADDED Software developers lead a stressful work-life, with long hours, strict deadlines, and many moving software components requiring constant coordination with multiple teams involved in the project.\ENDADDED}
Developers also transparently described  stressful working conditions specific to the organization culture. One developer from a multinational company working on a fully developed product discussed the ``aggressive focus on shipping elements'' and explains ``pretty tough to figure out something new to add to a mature product'' \citeVideo{M2_112}. Working in established companies with ``high pay means [a requirement for] a high accomplishment. It also means a high pressure'' which creates and accumulates over the years \citeVideo{M1_39}. At the same time, vlogs also described how the work culture of some organizations can be very supportive thus reducing stress: ``if you put your hand up the help will be there!'' \citeVideo{M1_34}.

\paragraph{Work-life Balance.} Working for long hours creates a disruptive work-life balance that vlogs were transparent about. 
To manage such full schedules, developers share different techniques such as creating a daily plan the night before:
``before bed I will take a look at my calendar and create a to-do list for the next day I write down two to three things that I need to get done in the next day and this routine helps me get ready for the coming day and also helped me narrow my focus on the most important things and then it's time for bed'' \citeVideo{M1_02}.
Several vlogs showed developers working late into the night  \citeVideo{M2_106}, \citeVideo{M2_121}, \citeVideo{M2_129}. ``After years of the stress for days and a night sitting long hours in front of the desks, it creates tension, headaches, low back pain, dry eyes, and obesity'' explains a senior developer \citeVideo{M1_39}, and this becomes a common topic of discussion among developers. In summary, the intention developers brought to how they experienced work stress and overcame it was a valuable part of a broader discussion on well-being.

\takeawayboxx{takeaway18}{\BEGINADDED Developers discuss various ways of dealing with stressful and large amounts of work to maintain work-life balance, deal with long term effects of such work on health and life.\ENDADDED}

\subsubsection{Lifestyle and Social Life}
\label{legacy:6.6}

\paragraph{Social activities and lifestyle.}
Developers in the vlogs were also intentional about showcasing that they engage in different social activities such as going out to restaurants and bars with friends \citeVideo{M1_69}, watching movies or playing video games with family and friends \citeVideo{M2_99}, \citeVideo{M3_135}, \citeVideo{M5_149}, and spending time with kids \citeVideo{M1_22}. Some developers engage in various service activities that intersect with their technical skills such as teaching high-school girls to code \citeVideo{M1_06}, organizing coding bootcamps \citeVideo{M1_33}, and arranging  club activities for underserved youth \citeVideo{M2_101}. Other developers show engaging in various creative hobbies like photography: ``So the other day I shot a engagement photo shoot and I need to edit those photos and videos'' \citeVideo{M1_17}.   

\takeawayboxx{takeaway19}{\BEGINADDED A large portion of the vlogs emphasize the life outside of work of developers. Developers discuss various hobbies and social activities they engage in after-work hours, as well as other healthy habits they build (and can afford) to lead a healthy lifestyle.\ENDADDED}

\paragraph{Healthy lifestyle and well-being.}
Some vlogs emphasized the importance of creating a healthy lifestyle that include regular workouts, and lead by example by showing different activities they engage in through the vlogs.
In these video segments, developers talked about the need to form habits to maintain in the longer run like staying hydrated \citeVideo{M1_57}, \citeVideo{M3_138}, practicing meditation \citeVideo{M1_42}. Many vlogs also focus on physical and social activities like sharing their workouts, healthy food recipes and eating habits to demonstrate how they maintain healthy lifestyle. Developers emphasize the importance of taking regular time out of the day to either take a walk, or go to the gym \citeVideo{M1_01}, \citeVideo{M1_26}, \citeVideo{M2_100}, or engage in various team sports with friends like tennis \citeVideo{M2_111}, basketball \citeVideo{M2_105}, and martial arts \citeVideo{M1_92}.

\medskip
The focus on lifestyle, health, and well-being not only breaks stereotypes painting developers as those who have an unhealthy with isolated life, but also provides a important information and examples for others in similar work situations.

\mybox{\faArrowCircleRight~\textbf{Summary:} \BEGINADDED Through the various topics discussed in the vlogs, developers share their experiences from learning programming and their career as a developer are different situations, to their journey as a developer in the long run and the effect these career decisions have on the overall quality of life.\ENDADDED}
\BEGINADDED 
\section{Study 3: \RQC (RQ3)}
\label{sec:study-comments}

Through Study 1 and 2, we find the motivations of the developers to post vlogs and the different topics developers discuss through the vlogs which not only includes information about their work-life but also how they manage their life outside work. However, how do the viewers receive these vlogs? What information do the viewers find useful and what kinds of interactions build around these vlogs? To answer these questions, we look into the comments section of each vlog.

\subsection{Approach}%
Unlike the number of views, which neither indicate the number of unique viewers~\cite{marciel2015understanding, keller2018flourishing}  nor their impression of the viewers who left comments on the videos had strong feelings (of agreement or disagreement) to interact through comments. While it is not feasible to interview all the viewers as they are both unidentifiable and numerous, analyzing the comments helps us infer who these viewers are and how they perceived information within the vlog. We use the comments on the vlogs as a form of impact, since the comments capture the communication that happens around a vlog and also how the vlog is perceived. The vlogs generated a lot of interaction through comments, with an average of 516.22 comments across 130 videos (min. no. of comments on a vlog = 0, max. no. of comments on a vlog = 13,522).

\paragraph{Data Collection.} To understand the type of conversations and community these vlogs generate, we collected a total of 1176 comments from the 130 videos. We selected the first 10 comments from each vlog (for videos with $<10$ comments, we collected the all the comments). Upon initial analysis we found that beyond the top 10 comments, was not useful nor vastly different from the top 10 comments. In addition, the number of interactions with the comments (likes, replies etc.) decreased heavily after 10 comments. We excluded pinned comments from the vloggers themselves, which typically summarized the video or linked to the creators' other social media platforms or text to promote of sponsored products. Each comment for the particular video is referred throughout the paper in the format [<video ID>-<comment number>] e.g. the $8^{th}$ comment for the video \citeVideo{M2_107} is referred in the paper as \citeComment{M2_107}{8}.

\paragraph{Analysis.} The first author categorized the comments looking for two types of information. First, the community where the viewer belonged e.g. ``I am a vlogger who also made \ldots'' we identified the viewer as a developer from the vlogger community, ``I am a student in computer science \ldots'' as a student, ``I am a Software Engineer in India and my days are \ldots'' as another developer. Other comments which didn't directly suggest the community we categorized as unidentified.

Second, we analyzed the type of comment using an inductive open coding approach~\cite{Saldana2009} where the first author assigned descriptive codes (a short phrase or label) to each comment based on how the viewers found value in the video and the type of video e.g. sharing their gratitude, asking technical questions. We identified $15$ such codes. The first and second author collaboratively, through negotiated agreement, split and merged codes to finally have $20$ codes. Then, they performed axial coding to group similar codes into $5$ high-level categories that summarize the intention and information within the comments. %
\ENDADDED

\subsection{Results}%
\label{sec:Comments}

To understand the impact, we look at the comments left by the viewers on the vlogs.
From our thematic analysis we find that these comments are from people who are from different walks of life. Through self-reported experiences we can identify some comments coming from developers, students, and those transitioning from other professions (e.g., nurse, sales industry, etc.)
With a focus on understanding what the audience finds valuable in the vlogs and infer what impact it has on the larger community, we analysed the 1176 comments to identify perceived value. In our supplementary materials, we include our codebook describing the categories of analyzed comments, the specific codes within each category, and an example of each.

\smallskip
\subsubsection{Finding information}
\label{legacy:7.1}

Commenters found the overall picture of a developers' life captured in the vlogs beyond what is described in job descriptions and articles. The ``talks about everyday developer challenges'' \citeComment{M125}{5} and ``actual look into the workflow of others'' \citeComment{M54}{3} paints a complete picture.

Commenters described how vlogs provide ``valuable insights and suggestions'' like tips on interviewing and making resume \citeComment{M24}{2}, programming within specific domains like ``foundation knowledge on Shopify themes'' \citeComment{M54}{5}, or usage of particular commands like ``how setTimeout is used in the real world'' \citeComment{M54}{10} or  ``glimpse of [how] some of the technology she used in the video like Jira'' \citeComment{M125}{5}. Further, commenters also found lifestyle related information valuable. A fellow developer stated they found one vlogs ``very informative'' as he too ``may be moving to Chicago for work very soon, and the video really helped [him] learn more about the city.'' \citeComment{M95}{4} Developers from around the world---Morocco \citeComment{M106}{2}, Dublin \citeComment{M113}{5}, USA \citeComment{M128}{3}---can find information about the life of developers vlogging from other parts of the world like China \citeVideo{M2_113}, Japan \citeVideo{M1_69}, and Philippines \citeVideo{M1_10}. Not only region, commenters also find information about development within specific industries and work settings like ``what freelancing would be like'' \citeComment{M22}{7} or ``working for a financial tech company'' \citeComment{M95}{6} within the vlogs.

\takeawayboxx{takeaway20}{\BEGINADDED Viewers find value in vlogs focusing on developer lifestyle as well as those sharing work related information like specific tools/languages. These vlogs eventually encourage viewers to consider pursuing a career in development.\ENDADDED}

This holistic image of developers captured in the vlogs inspire others to pursue development as a career, breaking boundaries of stereotypes through information and awareness: ``Because of [the vlog] , I'll pursue my dreams'' \citeComment{M50}{1} The personalized view of the developers through vlogs feels like getting to know another developer and leaves many commenters saying ``I wanna be a software engineer.'' \citeComment{MS7}{1} Commenters thank the vloggers for information the vlogs provide and encourage them to continue posting content online.

\smallskip
\subsubsection{Seeking advice}%
\label{legacy:7.2}

When commenters discovered information they care about in the vlogs, they followed-up with questions seeking clarification or further insights through the comments. Clarifying their own interpretations and seeking further information can reduce misconceptions about the identity and background requirements of being a developer. Some of these clarifications were in the form of direct questions about getting started with programming: ``Where to practice programming/coding?'' \citeComment{M33}{9} ``After learning Python how to start freelancing?'' \citeComment{M63}{5} While others were questions about particular tools and languages like ``Which language and framework are you working on?'' \citeComment{MS17}{9} or ``What DevOps tools did you use?'' \citeComment{M106}{8} Questions about salary \citeComment{M30}{4}, \citeComment{M42}{4}, \citeComment{M150}{7} and living conditions like apartment \citeComment{M146}{1} \citeComment{MS2}{4} or office facilities \citeComment{M11}{10} and working hours \citeComment{M145}{10} was also common among the comments. 

Commenters often provided some additional context about their current situations to get personalized advice regarding career or education. For example, one commenter provided information about his country and educational background in hopes of getting tailored advice: ``Can you help me to know something? Is having a computer science degree essential to get a web developer job in India? I have intermediate level understanding in front end development \ldots but do not have any degree in computer science.'' \citeComment{M21}{5} Commenters, especially those who explicitly identified themselves as students, often seek such career related advice on getting a degree for specific research interests \citeComment{M55}{1}. While other commenters seek information about changing careers \citeComment{M68}{2}, \citeComment{M71}{5} and moving from full time to freelance. Once in a while, commenters requested vloggers to post specific types of content that would directly benefit them: ``Could you please make a video about how did you get a job in Korea?''  \citeComment{M135}{1}

\subsubsection{Expressing empathy}%
\label{legacy:7.3}

Vlogs, through their unique personalized narratives, invoked feelings of empathy among some viewers who left comments exclaiming how they could relate to situations and struggles captured in the vlogs.
Commenters praised the vlog's ability to portray the realistic sides of being a developer that often contradict popular beliefs and stereotypes and software developers: ``This video goes on to show a very important point - software engineering is not equal to just coding. It is about human interactions to come to a decision and then execute that decision (of which a part is coding other part may be documentation or operations).'' \citeComment{M5}{6} Commenters often resonated with the reality of developers' life captured, one that's filled with hectic schedules and hard work: ``I wake up at 7:00 be at work at 8:30, start fixing bugs until 6:00 pm, be at home at 7:30.'' \citeComment{M14}{9}

Commenters who relate to the vlogs often share details about their own life. While some shared their own schedules and workflows, others shared their journey of learning programming: ``I was 14 when I started programming and now I am 27 years “old” and I am still learning new things on a daily basis (both intentionally and unintentionally).'' \citeComment{M78}{8} These experiences through comments provide multiple perspectives of \emph{who a developer is} and \emph{what they do} to other viewers who might come across the comments section.

\takeawayboxx{takeaway22}{\BEGINADDED Viewers found vlogs relatable to their own work and life experience, and in return would share their stories and experiences in the comments. These interactions provided multiple perspectives of the ``developer experience'' that even other viewers can access.\ENDADDED}

\subsubsection{Discovering a community}%
\label{legacy:7.4}

We also found that vlogs and their comments can be used as a tool to discover a community and celebrate each other. For instance, one commenter in particular was proud to see another person who looked like them they could relate to:  ``I feel so proud to see black women in IT and Engineering field. I am also a database administrator. Proud of you girl!'' \citeComment{M6}{3} Communities in the comments formed around race, region, religion, and other intersections of personal identities vloggers shared with their audience. For example, one commenter who identified themselves as a software engineer and a mother described how encouraged they were to see someone just like them, ``I’m so happy to find this video!! I am also a mom software engineer \& can't wait to get this show on the road!'' \citeComment{M22}{3}  Likewise, commenters also left encouraging comments for other viewers such as, ``I am a disabled Veteran. It's never too late. Coding has changed my life.'' \citeComment{M78}{1} Vloggers sharing their intersecting personal identities while sharing encouraging stories about their professional experiences motivate commenters to do the same amongst themselves thus fostering a community.

We also found that communities formed around the profession of being a vlogger in itself. One way this occurred is when a groups of commenters who are familiar with a particular vlogger's content bond and share jokes: ``Sweet baby Jesus, a regular upload schedule. I must be dreaming!'' \citeComment{M74}{3} %
Another way we noticed a community forming around this profession is when other vloggers who are software developers leave encouraging messages and feedback  through the comments such as, ``Great video [vlogger's name], many people will find this helpful!'' \citeComment{M47}{9} \BEGINADDED The vloggers also participated in discussions and responded to comments. \ENDADDED The several types of communities that formed around these vlogs created a sense of belonging and provided a platform to inspire each other.

\takeawayboxx{takeaway23}{\BEGINADDED The conversations around the vlogs eventually lead to various communities. Viewers find and encourage the creators and other viewers who leave comments who are in similar work or life situations.\ENDADDED}

\subsubsection{Dissent and Sarcasm}
\label{legacy:7.5}

While some commenters agree and empathized with the content shared in vlogs, we also had a group of comments that did where some shared dissent around what was included or shared satire jokes about developer experiences. 

Some comments sarcastically described the vlog as skewing the image about a developer's life portrayed in the vlogs: ``Seems like a great life to me, working out, good lunch, diner and dessert, playing pool, joining a developer conference with free pizza \ldots oh and of course a little bit of coding besides the important stuff.''~\citeComment{M92}{7} The differing opinions about developer life from comments further highlights the diversity in activities and identities among the developers. %
Another set of comments were about making parodical statements about programming in general like ``I usually do programming under water, so nobody can see me cry'' \citeComment{M138}{1} or about the life of developers like ``No hugs and kisses. Only bugs and fixes - Life of a programmer \citeComment{MS8}{5}.''

\medskip

Our findings from our comment analysis suggest that vlogs were able to reach a broad audience and those who commented were able to find valuable information in a community that supported a variety of perspectives of software development. 
This cumulative impact of vlogs from developers around the world, with diverse backgrounds and circumstances, can contribute to a shift in expectations of what a software developer's life includes. 

\mybox{\faArrowCircleRight~\textbf{Summary:} \BEGINADDED Commenters used vlogs to find information related to work and lifestyle of software developers, seek advice about technology and careers, express empathy, and to discover a community of people they can relate to.\ENDADDED}

\section{Discussion}
\label{sec:discussion}

From our three studies, we find that the intersection of vlog creators' intentions, vlog content included, and engaged vlog commenters help reinforce a community around each video. Here we summarize the main takeaways from our findings:

\BEGINADDEDX
\emph{(\textbf{RQ1}) \RQA} We found that developers aim to show the ordinarity of their daily lives to debunk the perception of general public that software workers are specially intelligent and devotes all of their life to work. Such a perception might discourage people from considering a potential career in the computing industry. Developers emphasize how their job is much like other professions, where they have a social life and hobbies outside of work, and that programming and other activities related to programming are skills they acquire over time through practice and learning. To encourage marginalized communities in programming related careers, developers share specific experiences as a way to provide helpful resources and a supportive community. Section~\ref{sec:rq1_res} highlights other motivations behind developers creating vlogs in Findings~\ref{takeaway1}-\ref{takeaway8}.

\emph{(\textbf{RQ2}) \RQB} Driven by these motivations, developers emphasize specific aspects of their life in these vlogs. To create awareness and encourage more people, developers discuss their own journey of learning how to code and share other technical experiences and skills important for developers, e.g., collaboration and communication. Developers also discuss various pathways to learn and work in the software industry today as well as the pros and cons of self-learning, freelance, remote work, etc. Outside of work, developers share information about their lifestyle, living conditions and hobbies they can afford, after-work routines, and share how they pay attention to their health and deal with common health complaints in the industry like stress or fitness. Other types of content that are commonly found in the vlogs are discussed in Section~\ref{sec:Topics}, highlighted in Findings~\ref{takeaway10}-\ref{takeaway19}.

\emph{(\textbf{RQ3}) \RQC} The information shared through vlogs are informative and perceived useful by many of the viewers, as they leave comments on the videos expressing their gratitude or opinions. These conversations often lead to small communities where the developer who vlogged and their audience encourage each other with shared experiences. Additionally, viewers add their own experiences when facing similar situations described in vlogs. These comment interactions create an archive of different perspectives, settings, and lifestyle related experiences for other audiences who come across the vlog to engage in. Other ways the community of viewers receive these vlogs are described in Section~\ref{sec:Comments}, highlighted in Findings~\ref{takeaway20}-\ref{takeaway23}.
\ENDADDED

Next, we discuss how these vlogs have emerged as a way to combat stereotypes of software developers, provide recommendations for future vloggers, and implications on how vlogs can be used in research.

\subsection{Stereotypes in Computing}
\label{sec:stereotypes}

Dismantling stereotypes around developers emerged as the higher-level theme throughout Studies 1, 2, and 3. The stereotypes are informed by the vloggers' motivations, topics discussed in the videos, and the comments from viewers.
From these studies, we identified three categories of 10 stereotypes dismantled through multiple rounds of organizing and re-grouping them until categories were distinct. 
Table~\ref{tbl:stereotypes} shows the 10 stereotypes distributed across categories. This table also includes references to subsections in the paper (listed in parentheses) and a short list of interview participants and vlogs analyzed that addressed each stereotype. The data sources is not exhaustive list of all evidence from our analyses. Each stereotype we present includes evidence from the interviews, vlog content analysis, and vlog comments.

\begin{table*}[!t]
\caption{\BEGINADDED Stereotypes dismantled about software developers from our study.\ENDADDED}

\label{tbl:stereotypes}

\begin{tabular}{lp{12cm}}
\toprule
& Stereotype / Links to Evidence \\
\midrule
& \bf\textsc{Who is/can be a developer} \\[3pt]
S1 & Developers are mostly male and mostly white (of European descent) \\ 
\small & \small $\rightarrow$ Sections~\ref{legacy:5.1}, \ref{legacy:6.3}, \ref{legacy:7.1}, \ref{legacy:7.2}, \ref{legacy:7.4} \hfill [\video{M6}, \video{M100}, \participant{V16}, \participant{V21}, \ytComment{M6}{8}, \ytComment{M101}{2}, \ytComment{M135}{6}] \\[3pt]

S2 & Developers are a young crowd, with no responsibility other than themselves \\
\small & \small $\rightarrow$ Sections \ref{legacy:5.3}, \ref{legacy:7.4} \hfill [\video{M33}, \video{M71}, \video{M22}, \participant{V04}, \ytComment{M118}{6}, \ytComment{M78}{7}] \\[3pt]

S3 & Developers are math wizards and they are born with coding skills \\
\small & \small $\rightarrow$ Sections \ref{legacy:5.1}, \ref{legacy:6.1}, \ref{legacy:6.3}, \ref{legacy:7.2} \hfill [\video{M47}, \video{M82}, \participant{V16}, \participant{V20}, \ytComment{M152}{9}, \ytComment{M3}{10}] \\[3pt]

S4 & Getting a traditional CS degree is essential to be a developer \\
\small & \small $\rightarrow$ Sections \ref{legacy:5.4}, \ref{legacy:6.1}, \ref{legacy:7.2} \hfill [\video{M3}, \video{M9}, \video{M100}, \video{M102}, \participant{V17}, \ytComment{M78}{7}, \ytComment{MS_19}{5}] \\

\midrule
& \bf\textsc{Work life of developers} \\[3pt]
S5 & \strut\BEGINADDED Developers code all day and knows nothing beyond it \\
\small & \small $\rightarrow$ \BEGINADDED Sections \ref{legacy:5.3}, \ref{legacy:6.2}, \ref{legacy:6.5}, \ref{legacy:7.3} \hfill [\video{M19}, \video{M48},  \video{M49}, \video{M99}, \participant{V04},  \participant{V16},  \ytComment{M102}{7}, \ytComment{M118}{1}, \ytComment{M138}{6}] \ENDADDED \\[3pt]

S6 & Developers seldom talk to others \\
\small & \small $\rightarrow$ Sections \ref{legacy:6.2}, \ref{legacy:7.3} \hfill [\video{M29}, \video{M85}, \participant{V06}, \participant{V08}, \ytComment{M24}{5}, \ytComment{M5}{6}] \\[3pt]

S7 & Stereotypes about job titles, startups, freelancing, and organizations \\
\small & \small $\rightarrow$ Sections \ref{legacy:6.3}, \ref{legacy:6.4} \hfill [\video{M23}, \video{M9}, \participant{V08}, \participant{V14}, \participant{V06}, \ytComment{M23}{10}, \ytComment{M30}{6}] \\

\midrule
& \bf\textsc{Personal life of developers} \\[3pt]
S8 & Developers have no time for fun \\
\small & \small $\rightarrow$ Section \ref{legacy:6.6} \hfill [\video{M2}, \video{M135}, \participant{V08}, \participant{V12}, \participant{V05}, \ytComment{M4}{2}, \ytComment{M102}{8}] \\[3pt]

S9 & Developers are asocial or anti-social, and prefer to be left alone \\
\small & \small $\rightarrow$ Sections \ref{legacy:5.2}, \ref{legacy:6.6} \hfill [\video{M21}, \video{M31}, \video{M66}, \participant{V13}, \ytComment{M128}{8}, \ytComment{M67}{3}] \\[3pt]

S10 & Developers lead an unhealthy lifestyle \\
\small & \small $\rightarrow$ Sections \ref{legacy:6.5}, \ref{legacy:6.6} \hfill [\video{M13}, \video{M39}, \video{M149}, \participant{V02}, \participant{V22}, \participant{V23}, \ytComment{M14}{5}, \ytComment{M113}{1}] \\
\bottomrule
\end{tabular}
\end{table*}

\medskip

\subsubsection{Breaking Barriers of Stereotypes}

The motivations of developers who vlog along with the topics they discuss challenge misconceptions and stereotypes around developers work and life. These stereotypes eventually become perceptions and create barriers around the developer community, either consciously or unconsciously. We identified three themes of stereotypes that are addressed. \BEGINADDED The three themes of stereotypes dismantled through our studies are centered around who can be a developer, what the work life a developer includes, and what developer's personal life looks like.\ENDADDED

\paragraph{Who is/can be a developer} [S1-S4] These are stereotypes related to the identity of developers such as ``Developers are mostly male, and are of European descent'', ``Developers are a young crowd, with no responsibility other than themselves'', ``Developers are math wizards, and they are born with coding skills'', and ``Getting a traditional CS degree is essential to be a developer''. These stereotypes \textbf{create barriers for people of different backgrounds to embrace this community}, and be part of it. Developers from around the world, from all genders, are challenging the notion of the stereotypical image of male \BEGINADDED white/Caucasian \ENDADDED developers. The online presence of developers on \yt has made an impact in breaking this stereotype [Takeaway~\ref{takeaway5}] \participant{V16} talks about how ``the image that is currently on YouTube is that there are different kinds of people. It's not just one type of people, one type of race'' [Takeaway~\ref{takeaway6}]. Developers further challenges stereotypes in this theme by talking about having a family \citeVideo{M1_33} [Takeaway~\ref{takeaway14}] or sharing their experience as an older developer \citeVideo{M1_71} [Takeaway~\ref{takeaway15}]. Many developers discuss their educational background in their vlogs and talk about how they went to school for writing, statistics, or art and later switched their jobs by learning how to code through online courses and self-learning \citeVideo{M1_03}, \citeVideo{M1_47}, \citeVideo{M2_100}, \citeVideo{MS_19}. Developers also emphasize that math skills are not essential to be a developer and that they learned about coding only in college or even later: ``I'm not a genius, I don't know trigonometry or calculus or an advanced math. I went to school originally for creative writing. [\dots] I'm not great at math, I'm not a genius, I'm not a wizard. I'm not anything like that. [\dots] I was not a kid who hacked on computers all day'' \citeVideo{M1_47} [Takeaway~\ref{takeaway7}, \ref{takeaway10}].

\paragraph{Work life of developers} [S5-S7] Many stereotypes are related to how developers work: ``Developers code all day'', ``Developers are people addicted to coding and knows nothing beyond it'', and as a result ``Developers seldom talk to others''. In reality, coding is only a part of the daily activities [Takeaway~\ref{takeaway2}]. \BEGINADDED Our 3 studies dismantled this stereotype by showing that developers do much more than coding [Takeaway~\ref{takeaway1}]\ENDADDED. They need to learn and design the solution, communicate with designers and architects, and developers who work freelance need to apply for clients and handle their requests \citeVideo{M1_09}, \citeVideo{M1_22}. Developers spend a considerable amount of time at work collaborating with others, or in meetings or mentoring sessions[Takeaway~\ref{takeaway12}]. A developer spends ``a lot of time talking to others and thinking about work, while I'm on my editor for at most 2 hour'' \citeVideo{M2_99}. 
\BEGINADDED Vlogs show that in order to collaborate and mentor others successfully, its important to be a good communicator\citeParticipant{V06}.\ENDADDED

There are many stereotypes related to job titles, startups, freelancing, and organizations. In the videos, developers clarify the different positions available in industry (for example, front-end developer, back-end developer, full-stack developer, web developer, iOS developer) and emphasize the fact that no role is superior to another: ``I don't think that being a full stack developer is any better than being a front-end developer or a back-end developer it's just one different flavor and there's definitely drawbacks or and strengths strengths and weaknesses'' \citeVideo{M1_23}. \BEGINADDED Developers in our study combat the perception that only big companies are worth working for by emphasizing the advantages (and challenges) of working as a freelancer or for a start-up [Takeaway~\ref{takeaway14}, \ref{takeaway15}]\ENDADDED. For instance, ``small, empowered, well-funded startups are the ones that are game changers'' \citeParticipant{V06} and ``give you more freedom and impact'' \citeParticipant{V08}. The development world is ``constantly evolving, constantly changing'' \citeParticipant{V14} and embracing the different forms of the software development jobs beyond the traditional career will open new opportunities.

\paragraph{Personal life of developers} [S8-S10] Another group of stereotypes is related to the developers' personal life: ``Developers have no time for fun'', ``Developers are asocial or anti-social, and prefer to be left alone'', and ``Developers lead an unhealthy lifestyle''. 
\BEGINADDED The image of developers having no fun and being introvert and socially awkward is discouraging for those interested in entering (or staying in) the software development community and do not identify as introverts. Our studies clarify that developers can have a fun personal life with various hobbies and interest beyond work [Takeaway~\ref{takeaway3}, \ref{takeaway4}]. Our findings show that some developers spend time with family \citeParticipant{V04}, \citeVideo{M1_85}, while others enjoy activities with friends like sports, gaming, music, going to events and parties, painting and cooking \citeParticipant{V11}. \ENDADDED 
Many vlogs include activities showing a healthy lifestyle with workouts \citeParticipant{V05}, sports \citeVideo{M2_105}, healthy food, and health related practices such as hydrating and practicing meditation \citeVideo{M1_42} [Takeaway~\ref{takeaway18}, \ref{takeaway19}].
\BEGINADDED The stereotype of developers being ``willing to code 60, 70 hours a week and drink Red Bull and wear a hoodie'' \citeParticipant{V21} is diassembling thanks to vlogs:\ENDADDED
``there are more and more tech lifestyle YouTubers coming out who don't look like that, who don't live \emph{that} life'' \citeParticipant{V23}.

\BEGINADDED It is important for any community to be aware of stereotypes and how it can hinder the health and success of the community. In this paper, we showed how software developers' vlogs can dismantle and breaks these stereotypes. Besides vlogs, there are many other mechanisms that can be used to change stereotypes such as blogs and TikTok videos. A direction for future work is to study how these broader platforms are being used to address stereotypes, not just in isolation but also whether efforts are concerted across platforms.\ENDADDED

\BEGINADDED
\subsection{Vlogging for Developers}
\label{sec:developers}
\ENDADDED

\BEGINADDED

Vlogs are an important source of information for developers, not just about their work and careers, but also about other important topics such as work-life balance, stress management, and their personal lives in general.
Vlogs provide an effective way to build a community with people from various professions. Creating vlogs allows developers to showcase their technical and life-management skills, potentially creating opportunities in terms of job offer. Additionally, vlogging requires reflecting on how to effectively communicate and structure a story, which are valuable skills to have as a developer. Based on our findings, we discuss some strategies for developers to create vlogs.

\subsubsection{Best practices for creating vlogs}

When creating vlogs, just ``film[ing] as much as possible and then edit[ing] it out later, without really plan[ing] out anything, takes too much time'' \participant{V18}. Time is a valuable resource, as most developers use their after-work hours to create vlogs. From our results, we find six elements of a vlog which can guide designing the content of vlogs.

\begin{enumerate}
    \item \textit{Pick the theme or concept}: Identifying the motivation driving the vlogs early sets the tone for the remaining elements. Although, some developers prefer to vlog about instances when interesting things happen in their day-to-day life [\participant{V13}], it is difficult to continue producing content simply based on the chance of exciting events.
    From our findings, we identified five different motivations that drive the vlogs: \textit{Challenging misconceptions about developer identities} [Takeaways \ref{takeaway1}, \ref{takeaway2}, \ref{takeaway17}], \textit{Networking and building a community} [Takeaways \ref{takeaway3}, \ref{takeaway4}], \textit{Promoting diversity and minority communities} [Takeaways \ref{takeaway5}, \ref{takeaway6}], \textit{Creating awareness about computing as a career} [Takeaways \ref{takeaway7}, \ref{takeaway10}] %
    , or \textit{Supporting long term careers} [Takeaway \ref{takeaway14}]. %
    
    \item \textit{Define your audience}: Based on the theme, defining the target audience can help identify the depth of content and style of narration. When trying to reach a broad population, developers have to ``consider how easy the information [included in the vlog] is to digest'' [\participant{V22}], and reduce the use of ``jargon'' [\participant{V21}]. However, it is important to keep in mind that the vlogs are ``going to appeal mostly to a software engineering audience because of the nature of the topics talked about'' [\participant{V21}], hence all 130 vlogs included some technical content like meetings, or length of workday. The target audience can be defined using geographical boundaries [Takeaways~\ref{takeaway6}, \ref{takeaway8}], gender [Takeaway~\ref{takeaway5}], profession [Takeaway~\ref{takeaway10}], or age. %

    \item \textit{Choose a Genre:} Two main contrasting narration styles were prominent among 130 vlogs we analyzed. Following the traditional approach, vlogs can be created in poetic documentary mode~\cite{nichols2017introduction}: a chronological juxtaposition of short clips with no additional narration in time continuity (and maybe some background music), e.g., showing a clip of them driving to work, which cuts to them at work and setting up their laptop. Alternatively, developers used ``jump-cuts'' to ``cut out the boring parts'' keeping the viewers engaged throughout. As \participant{V12} explains, ``Pay attention to people's attention spans \ldots{} every second has to be something valuable.'' Occasionally these genres were mixed with comedic elements ``providing some level of entertainment value'' [\participant{V21}], which further engages the audience.   %
    
    \item \textit{Outline the narrative structure and tone:} The narrative structure and tone make a video's message resonate with its audience. When building the structure, \participant{V18} writes ``scripts, not line by line, but some broad strokes to talk about and then go from there.'' When writing such outline, \participant{V20} walks the audience through the story by ``first giving some foundation of current thought process, then taking [the audience] on the journey with me, reliving different aspects until the end of the story. It helps them put themselves into my shoes.'' Several developers set the narrative tone of their vlogs as ``authentic and real'' [\participant{V21}]. %
    This creates a feeling of relatableness and empathy with audience [Takeaway~\ref{takeaway22}].

    \item \textit{Select the content and editing style}: Based on the previous four elements, developers decide what specific topics to include, to what extent, and how? For example, when motivated to create awareness about developer careers among international students interested in studying in USA, \participant{V22} focuses on information that was difficult for him to find and discusses it ``to make sure that it's easy to find for somebody who is in that position.'' For such informative content, developers usually followed a lecture style edit with the aid of occasional lists of discussion items being discussed shown onscreen. When the content was more towards creating awareness about lifestyle and work-expectations from novice developers, \participant{V21} used a ``fresh'' editing style ``like word bubbles and zoom[ing]'' into transitions. The type of music/background sound effects used can further improve the appeal of vlogs to the audience. 

\end{enumerate}

Decisions made about these five elements gradually refine the idea of a vlog into production ready content. The earlier elements of the list act as filters for the later elements. Once the developer can identify the content and editing style (Step 5), the developer has a clearer idea about what to record and how.

\subsubsection{Design guidelines for platforms}

Based on our findings, we identified 10 stereotypes that developers dismantle using vlogs on \yt. These stereotypes [S1-S10] conforming developers to a specific identity act as barriers to diversity within the computing community. To help developers overcome these barriers, we discuss some opportunities to improve the video hosting platforms, especially \yt.

\begin{itemize}
    \item \emph{Highlighting marginalized communities.} A critical mass of developer vlogs are centered around promoting diversity by supporting underrepresented communities connecting with each other [Takeaway~\ref{takeaway5}, \ref{takeaway6}] and discovering useful resources [Takeaway~\ref{takeaway3}, \ref{takeaway10}]. However, it is difficult to identify these videos among other millions vlogs. \yt can improve the discoverability of these vlogs by introducing tags for videos. Currently, \yt (and Twitch) creators manually select from available tags (eg. \#developerlife)~\cite{twitchtags}. However, these tags are generic and not very informative (e.g, what about developer life? health? career opportunities?) and usage of overlapping or too many tags makes it more confusing for the viewers. Our findings on why people create or consume developer vlogs, and similar studies on other kinds of videos, can help create informative tags.
    
    \item \emph{Enable easy networking.} Developers found \yt as a platform that allows new ways of connecting with different people to spreading personal/social/online brands and for \participant{V08} this made \yt a \inlinequote{perfect window for making connections}. On \yt, viewers' urge to connect with creators is a spontaneous reaction after watching their content. This interest can be short-lived and without easy to access networking features, both the audience and creators miss out on potential connections. Currently, to reach out to a creator on \yt, viewers either have to leave comments and wait for the creator's response, or find out other communication channels such as email or Instagram to directly message them. Our sample of vlogs had an average of 516 comments making it hard for the creators to read them all. These asynchronous and external communication channels further hinder connectivity within communities [Takeaway~\ref{takeaway4}], and the platform can help provide features to help with these motivations rather than just being a broadcasting platform.

    \item \emph{Integrating viewer perspectives.} We find that comments generate a lot of interaction around vlogs with the creators as well as other viewers [Takeaway~\ref{takeaway23}, \ref{takeaway22}]. However, on a personal computer, the viewer has to scroll at least three times to read 6-8 comments. On mobile interface, the viewer has to intentionally expand the section. Further, when comments mention a specific part of the video the viewer has to scroll up to the video and find the correct timestamp to understand the context of the comment. These additional steps are obstacles to viewers attempting to quickly gather valuable perspectives from other developers and viewers from the comments [Takeaway~\ref{takeaway23}]. Platforms like SoundCloud~\cite{soundcloud-comments} allow users to see comments embedded on the player's progress bar which pop up as the video plays. Creating similar features for viewers to choose to see comments integrated with the content can support the motivations to create more awareness about ``who are developers'' [Takeaway~\ref{takeaway22}].

\end{itemize}

We should note that these best practices and guidelines apply to a range of roles outside of software development.
Further studies should be done on any type of video of knowledge workers, not just vlogs, to support the different motivations driving the videos, and communities that resurrect around them.

\subsection{Research Implications and Directions for Future Work}
\label{sec:research-implications}

From a research perspective, vlogs are a unique form of data that  researchers can use as a proxy for observational studies. Vlogs by nature provide an organic form of first-hand observation from the subject's perspective and not the observer. While this may hinder some inference, vlogs can serve as the time capsule for researchers to analyze how perspectives and societal beliefs surrounding software development have progressed over time. Vlogs also offer a unique pulse on how current events such as the COVID-19 pandemic and civil rights affect software developers and their life. In several vlogs, developers discussed how the pandemic affected their life \citeVideo{M2_125}, \citeVideo{M1_75} and led to cancelled projects \citeVideo{M2_127}. These vlogs could be used for example to study the collaboration challenges developer faced during the COVID-19 pandemic---a unique time period where many software developers were working remotely. 
In addition, we anticipate several other research opportunities related to vlogs: %
\ENDADDED

\begin{itemize}
    \BEGINADDED
    \item \textit{Lowering the barrier to vlog.} Creating vlogs about work is a common practice for software developers, however creating videos requires a significant effort and involves several steps such as recording, editing, and uploading files to YouTube. 
    A potential research direction is to focus on better tool support and the integration of  vlogging into software development processes. With the right tools, vlogs could attached to source code artifacts \ENDADDED to document important decisions that other developers could watch in the future while working. Given the transition to remote and hybrid work, this could also be an opportunity to increase the social connection among teams.
    \BEGINADDED

    \item \textit{Developer productivity and workday studies.} Researchers can use vlogs to further characterize workdays of software developers by focusing on the activities such as coding and meetings that are shown in vlogs. Previous workday studies have mainly focused on work activities~\cite{amann:saner:2016,ford2017characterizing,meyer2019today} but not on developer's life beyond work. Vlogs are a unique way to capture information in a non-intrusive way.
    Since vlogs also serve as a diary in a video form, they provide an excellent source to analyze how a software developer's life and work has changed over the years. How has software development changed over the past ten years? What work aspects have changed for developers? How have offices changed over the past decade?

    \item \textit{Social interactions at work (or in general).} Vlogs capture how developers communicate with coworkers as well as degree of engagement needed for different types of communications. Vlogs can be used to study how communication has changed while developers were working from home during the COVID-19 pandemic.%
    A related direction for future work are is to assess the impact of vlogging on productivity and team culture. While we found only two vlogs that featured more than one developer, creating a team vlog could be a valuable team building exercise.

    \item \textit{Beyond software development.} \ENDADDED
    The focus of this paper was on software developers and a future direction is to investigate how other knowledge workers such as accountants, lawyers, and doctors leverage vlogs. We expect that they face similar themes of stereotypes related to who they are and their work life and personal life. We also expect that they will employ similar strategies as software developers. For example, doctors may make and effort to clarify the different careers as doctors, show that the variety of work they do, and have a personal life beyond work. \BEGINADDED Vlogs can also be used to compare and contrast different job disciplines with each other.\ENDADDED
    
\end{itemize}

\medskip
Overall, vlogs are a tool of great value to all communities and can  influence a larger population by directly creating awareness through information through the lens of individual developers. Vlogs contribute in transforming the beliefs and perceptions of society to create a more accepting and diverse community of future generation developers.

\section{Limitations}
\label{sec:limitations}

Studies based on empirical data have their strengths and weaknesses. While field observation studies capture how participants act, it requires effort in collecting the data in presence of the researcher and the participant in their natural environment. In our case, we obtained this observation from existing vlogs on \yt that document developers in their own environment. In addition, vlogs capture two unique perspectives difficult to observe through shadowing, diary studies, or retrospective interviews --- (1) vlogs capture more than just the work; it captures their everyday activities like morning routines, social interactions, and other humane sides of developers that often overlooked; (2) vlogs contain information where the developers have control over the narrative, untainted by the researchers.

Consistency refers to ensuring that the results consistently follow from the data and there is no inference that cannot be supported after the data analysis. We established consistency in our findings by considering three sources of information -- talking to the the vloggers, analyzing the vlogs, and analyzing the comments from viewers. Through the interviews with the vloggers we triangulated the topics we identified through our interpretation of the vlogs to be consistent with the motivations behind the vlogs. And through the analysis of the comments we found that the commenters found value in the vlogs consistent with the vlogger's motivations and vlog's contents.

Vlogs on \yt have certain traits~\cite{cheng2013understanding} that conform to the general characteristics of videos on a popular entertainment platform like \yt  with over two billion daily users~\cite{youtubepress} e.g. average length approximately 11 minutes~\cite{clement2019youtube}, occasional elements of humor and suspense to keep the viewers engaged. These nuances of story telling often makes the vlogs more interesting and engaging for the viewers, and thus, it is further intriguing to study the effects vlogs have on dismantling stereotypes.

Establishing validity in qualitative research is challenging due to potential biases, including researcher bias, confirmation bias, and interpretive validity~\cite{Onwuegbuzie2007}. To reduce these issues, we categorized all data through several rounds of qualitative coding and thematic analysis across the three authors. Through negotiated agreement, the authors ensured the definition of the categories are independent, the categories are self-contained, and there are no overlapping-themes.

The focus of this study was to understand the concepts and context around the problem~\cite{tracy2010qualitative, merriam2015qualitative}, which in our case, was understanding how vlogs are currently used to convey information and dismantle stereotypes around computing. Our findings are not generalizeable to all vlogs by all developers; there can be other stereotypes addressed by vlogs and technical videos beyond our study scope. Our findings only speak to the existence of the vloggers' efforts to dismantle stereotypes. However, we ensured that our findings are derived from data is diverse and representative through several study design decisions. First, we selected 130 videos from 113 developers spanning 21 countries from North America, Europe, and Asia to ensure are findings are not limited by social norms and stigmas from a few countries only. Second, our interview participants come from eight countries across three continents. They also worked in different size and type of organizations, positions of different titles like ``senior developer'' or ''full-stack engineer''. Although we couldn't identify the origin of the users for the 1176 comments, nor the people behind millions of views on these vlogs, we did not exclude spam or malicious comments to ensure our data set is not skewed to reflect only one extreme of the commenters' perception about the vlogs.

\section{Conclusion}
\label{sec:conclusion}

Through vlogs, developers showcase themselves as more than nerds who work behind computers all day.
They share technical knowledge and professional experiences as well as rare moments of their personal lives that resonate with their viewers and inform them at the same time.
We found that vlogs were a means to help developers breakdown the misconceptions and pessimistic perspectives to a career in software development. 
Through vlogs, developers advocate for a diverse community of developers from different educational backgrounds and professional journeys.
From the comments we found that these vlogs provided a community of support where people bonded over shared experiences breaking the mold of what a software developer can be.
We discuss the types of perspectives developers share to address stereotypes around their identity, work, and personal life. \BEGINADDED We also outline est practices and implications for content creators, video sharing platforms, and research. \ENDADDED
In studying mechanisms where software developers control the narrative of their profession, we learn more about how technical workers use online platforms to challenge stereotypes and inspire a diverse generation of knowledge workers.

\bibliographystyle{ACM-Reference-Format}
\bibliography{acmart.bib}

\received{October 2020}
\received[revised]{April 2021}
\received[accepted]{July 2021}

\end{document}